%% file: new-new-2__31_.tex
\documentclass[12pt]{article}
\input{dstyle2}
\usetikzlibrary{calc}
\usetikzlibrary{matrix}
\usetikzlibrary{positioning}
\usetikzlibrary{decorations.pathreplacing}
\allowdisplaybreaks

\usepackage{etoolbox}
\patchcmd{\quote}{\rightmargin}{\leftmargin 4em \rightmargin}{}{}

\makeatletter
\AddToHook{cmd/appendix/before}{\def\cref@section@alias{appendix}\def\cref@subsection@alias{appendix}}
\makeatother

\begin{document}	
\title{Mediated Trust\thanks{I am indebted to Johannes H\"orner, Larry Samuelson, and Marina Halac for their support. I also thank Charles Angelucci, Dan Barron, Joyee Deb, Eddie Dekel, Francesc Dilm\'e, Mehmet Ekmekci, Jeff Ely, Françoise Forges, George Georgiadis, Som Ghosh, Michihiro Kandori, Jin Li, Bart Lipman, Elliot Lipnowski, Simon Loertscher, Lucas Maestri, George Mailath, Marco Ottaviani, Harry Pei, Andrew Postlewaite, Mike Powell, John Quah, Daniel Quint, Wojciech Olszewski, Guillermo Ordo\~nez, Juan Ortner, David Rahman, \href{https://refine.ink/}{Refine.ink}, J\'er\^ome Renault, Guillaume Roger, Anna Sanktjohanser, Johannes Schneider, Vasiliki Skreta, Alex Smolin, Takuo Sugaya, Satoru Takahashi, Juuso Toikka,  Juuso V\"alim\"aki, Rakesh Vohra, Asher Wolinsky, Yu Fu Wong, and audiences at various seminars for helpful comments. This paper supersedes an earlier version circulated under the title ``Dynamic Correlation as an Incentive Device.''}}
\author{Allen Vong\thanks{National University of Singapore, \href{mailto:allenv@nus.edu.sg}{\texttt{allenv@nus.edu.sg}}.}
}
\date{\today}
\hypersetup{pageanchor=false}
\maketitle

\begin{abstract}
I study how mediation facilitates trade between a long-lived seller and short-lived buyers. Without mediation, the seller's moral hazard bounds equilibrium payoffs away from his favorable part, and possibly all, of the efficiency frontier of the set of feasible and individually rational payoffs for the seller and the average buyer, even as he becomes arbitrarily patient. In contrast, with mediation, all interior payoffs in this set can be attained in equilibrium as the seller becomes patient. Mediation achieves these gains by privately informing the seller which continuation applies. Realizing these gains in full requires nonrecursive mediation.
\smallskip


\end{abstract}

\thispagestyle{empty}
\pagebreak
\setcounter{page}{1}
\hypersetup{pageanchor=true}

\section{Introduction}\label{sec:introduction}

Since the seminal work of \citet{klein1981role}, economists have studied how the prospect of future business disciplines a seller's moral hazard. In the now-canonical setting of a long-lived seller facing successive short-lived buyers, \citet{fudenberg1994efficiency} show that this discipline is imperfect: gains from trade are constrained.

Institutions that mediate transactions may promote gains from trade. \citet{milgrom1990role} illustrate this in their study of the Law Merchant. Today, online platforms play a similar role. The literature has focused on {\it public} mediation, such as ratings (\egg \citealp{ekmekci2011sustainable}; \citealp{horner2021motivational}; \citealp{vong2025rep}). Yet institutions also mediate {\it privately}. Online platforms such as Amazon and Walmart Marketplace inform each seller, through account dashboards and notifications that buyers cannot access, of his standing and its consequences, up to account suspension.\footnote{Amazon Seller Central, \href{https://sellercentral.amazon.com/gp/help/external/GR786P4BPEVKTBAG?locale=en-US}{``Account health rating''}; Walmart Marketplace, \href{https://marketplacelearn.walmart.com/guides/Policies\%20\%26\%20standards/Performance/Seller-performance-standards}{``Seller Performance Standards: Overview.''} Accessed October 19, 2026.}

I study how private communication improves gains from trade under complete information. In each period, a new short-lived buyer decides whether to trade with the long-lived seller and, if so, which of the available trades to undertake. These trades differ in their terms of exchange, such as quantity, price, or product grade. If she trades, the seller privately chooses an effort level, which then produces a good or bad output, with higher effort making a good output more likely. For each available trade, the buyer prefers to trade only if the seller exerts high enough effort; the seller prefers trade and finds higher effort more costly. Consequently, the seller's incentives to exert higher effort are intrinsically dynamic. The trade outcome---the buyer's action and output (or lack thereof)---is observable to the seller and future buyers. A mediator commits ex ante to a communication device that sends private messages to the seller and the buyers based on past messages, buyer actions, and outputs (\citealp{forges1986approach}; \citealp{myerson1986multistage}). While messages are private, communication can be public in any given period: the mediator can inform each player of the other's message. In fact, for my results, only the mediator's communication with the seller needs to be private.

Without mediation, the seller's moral hazard keeps equilibrium payoffs of the seller and the average buyer bounded away from the seller-favorable part, and possibly all, of the (Pareto-)efficiency frontier of feasible and individually rational payoff vectors, even if the seller is arbitrarily patient. This reflects the limitation identified by \citet{fudenberg1994efficiency}: even when patience makes dynamic incentives most effective, the seller's rewards for good outputs and thus his effort incentives are constrained by the buyers' myopic willingness to trade. The folk theorem of \citet{fudenberg1994folk} does not apply.

My main result overturns this  conclusion. With mediation, every feasible and individually rational interior payoff vector for a sufficiently patient seller and the average buyer can be attained in a strong sequential communication equilibrium (SSCE, \citealp{gerardi2007sequential}). Mediation can therefore virtually implement any Pareto-efficient payoff vector, including one that maximizes the average buyer's feasible payoff and another that gives the seller his mixed-action Stackelberg payoff.\footnote{This result corrects a claim in an earlier version circulated under the title ``Dynamic Correlation as an Incentive Device'' that the Stackelberg payoff cannot be approached in the patient limit.}

For each interior feasible and individually rational payoff vector, I construct an SSCE that attains it when the seller is sufficiently patient. The gains from mediation arise because private communication enables dynamic correlation: the mediator randomizes over private messages linked to different continuations, informing the seller but not the buyers which continuation applies. My construction uses finite ladders of trade states: the mediator informs the seller, but not the buyers, of the state, with higher states prescribing weakly lower effort. Buyers are willing to trade given the resulting distribution over efforts, while the seller strictly prefers the effort prescribed in each continuation. Continuations that prescribe low effort serve as rewards for good outputs and relax the constraints on effort incentives relative to the no-mediation benchmark. This is possible because mediation bypasses the indifference required for the seller to mix: without mediation, the seller's mixing over effort levels requires him to be indifferent among them, and this indifference constrains payoffs. The correlation must be dynamic: if the mediator induced a nontrivial distribution over the seller's efforts but made future play independent of messages---so that correlation is static rather than dynamic---the seller would still have to be indifferent among the efforts prescribed by that distribution, which constrains payoffs just as mixed strategies do.

In fact, \emph{any} payoff gains from mediation in my model require private communication that leaves buyers uncertain about which continuation applies. Because monitoring is public---buyer actions and outputs are publicly observable (and public randomization is available)---mediation whose messages are publicly observable does not expand equilibrium payoffs: its messages necessarily follow distributions determined by the public histories and can be replicated by public randomization. This public monitoring also implies that the gains from mediation rely neither on censoring buyers' information about the past (\egg \citealp{ekmekci2011sustainable};  \citealp{bhaskar2019community}; \citealp{best2024persuasion}; \citealp{pei2024reputation}; \citealp{vong2025rep}) nor on withholding or delaying performance feedback (\egg \citealp{abreu1991information}; \citealp{sugaya2025nonrecursive}). 

The public monitoring also highlights that the efficiency gains my results identify are robust to reduced observability. In practice, buyers may observe only recent trade outcomes. Such limited observability only strengthens the case for mediation. Without mediation, if buyers observe only recent outcomes, the equilibrium payoff set weakly shrinks, for a reason similar to that in \citet{kandori1992use}: the same equilibrium payoffs can be attained when buyers observe more, using strategies that ignore the additional information. With mediation, the mediator can disclose the full public history alongside its original private messages and implement the same outcomes as when buyers observe the full public history, so every interior feasible and individually rational payoff vector remains attainable when the seller is sufficiently patient.

From a technical perspective, public monitoring imposes a filtering constraint: the mediator must track each buyer's beliefs about the seller's private history, and hence his effort, given the public history and her private message. This prevents an application of methods for blind repeated games in which only the mediator observes outputs \citep{sugaya2017bounding, sugaya2023monitoring, sugaya2025nonrecursive} and calls for equilibrium constructions that keep the filtering tractable. In the SSCEs I construct, in periods with trade, buyers' beliefs about the seller's effort after receiving the mediator's message remain unchanged even as the public history evolves.

The SSCEs I construct are nonrecursive: some continuations are not SSCEs on their own. Indeed, achieving the efficiency gains in full requires nonrecursive mediation. I show that under perfect communication equilibrium (PCE, \citealp{tomala2009perfect}), which has a recursive structure, equilibrium payoffs remain bounded away from the seller-favorable part, and possibly all, of the efficiency frontier even as the seller becomes arbitrarily patient. This is because recursive mediation allows the mediator to reward good outputs only with continuations that are themselves equilibria, which limits the seller's effort incentives and, in turn, equilibrium payoffs. This contrasts with \citet{tomala2009perfect} in which a folk theorem holds under PCE and with \citet{sugaya2025nonrecursive} in which, under their identifiability conditions, the gains from nonrecursive over recursive mediation vanish in the patient limit.

A well-known alternative remedy for the payoff loss in the no-mediation benchmark is reputation: under suitable identification conditions, incomplete information about the seller's type---even a slight possibility that the seller is committed to effort with sufficiently high probability---promotes gains from trade and lets a patient strategic seller secure a high payoff \citep{fudenberg1992}. Recent work on reputation also studies the short-lived players' welfare (\citealp{pei2024reputation,pei2026reputation}). The identification conditions fail in my model: a buyer who expects shirking prefers rejection, which reveals nothing about the effort the seller would choose if trade occurred. Consequently, even if buyers entertained the possibility of a committed seller type, the reputation payoff lower bound of \citet{fudenberg1992} would be vacuous. And even in settings where the identification conditions hold, buyers may have no reason to believe that the seller could be committed. My main result implies that, without type uncertainty, mediation can bring either the seller's payoff arbitrarily close to his mixed-action Stackelberg payoff or the average buyer's payoff arbitrarily close to its feasible maximum, provided that the seller is sufficiently patient.

\paragraph{Related literature.} My results contribute to the literature on mediation. The gains from mediation that my results identify---eliminating the inefficiency that arises when buyers' myopic willingness to trade constrains the seller's dynamic incentives---differ from the known advantages of mediation, such as improving identifiability of hidden actions (\citealp{rahman2010mediated}; \citealp{rahman2012but, rahman2014power}), creating strategic uncertainty among long-lived players through private recommendations (\citealp{sugaya2018maintaining}; \citealp{ortner2024mediated}; \citealp{kawai2026value}), or eliciting private information in bargaining \citep{fanning2023optimal}. Here there is no action-identifiability problem, one long-lived player, and no private information to elicit. Moreover, while mediated strategies are private strategies, the gains of mediation in my result differ from those of private strategies in unmediated repeated games (\citealp{lehrer1991internal}; \citealp{mailath2002private}; \citealp{kandori2006efficiency}): here, unmediated private strategies do not improve payoffs over public ones as monitoring has a product structure \citep{fudenberg1994efficiency}.

My results relate to work on reputation without commitment types (\citealp{board2013reputation}; \citealp{bohren2024persistence}), where the seller's action affects a persistent payoff-relevant state. Here, the seller's action affects only current output; its persistent effects are designed by the mediator. \citet{pei2021trust} studies a trust game with type uncertainty but no commitment types, in which every seller type's highest equilibrium payoff remains below his mixed-action Stackelberg payoff even as he becomes arbitrarily patient. Here, mediation allows the seller to approach this payoff without type uncertainty. Information design in reputation models with commitment types (\citealp{ekmekci2011sustainable};  \citealp{vong2025rep}) relies on type uncertainty and on public communication that censors the public history; the gains from mediation here require neither.

\section{Inefficiencies without mediation}
\label{sec:benchmark}

In this section, I characterize equilibrium payoffs in a benchmark model without mediation. Time $t=0,1,\dots$ is discrete and the horizon is infinite. A long-lived seller, with discount factor $\gd \in (0,1)$, interacts with a sequence of short-lived buyers, each of whom lives for one period.

In each period, the buyer chooses an action $b\in B:=\{0,1,\dots,\overline b\}$, where $\overline b \ge 1$. Action $0$ means that the buyer rejects trade. Each action $b>0$ specifies a trade with given terms of exchange, such as quantity, price, or product grade.\footnote{Following the literature (\egg \citealp{mailath2001}; \citealp{ekmekci2011sustainable}; \citealp{levine2021reputation}; \citealp{pei2021trust}), I take these terms as given and let the buyer choose among the available trades.} The seller observes this choice. Upon rejection, the period ends. If the buyer trades, the seller privately chooses effort $a\in A:=\{0,1,\dots,\overline a\}$ to carry out that trade, with $\overline a \ge 1$; I refer to effort $0$ as shirking. The output is good, denoted by $\overline q$, with probability $p_{ab}\in(0,1)$ and is bad, denoted by $\underline q$, otherwise. Both the seller and the current buyer observe this output. Future buyers observe the current buyer's action and the output or lack thereof.

A public randomization device, whose realizations are drawn uniformly and independently across periods from the unit interval at the beginning of each period before the buyer moves, is available. It eases the exposition in this section and plays no role in my main result. As is customary, this device is omitted from the notation.

In each period, if the buyer rejects trade, both players receive zero payoff. If she chooses trade $b>0$ instead, the seller's payoff from effort $a$ is $u_{ab} := w_b-c_{ab}$, where $w_b$ denotes his benefit from the trade and $c_{ab}$ denotes his effort cost. This cost is strictly increasing and (weakly) convex in effort. Shirking---effort $0$---entails zero cost, $c_{0b}=0$. Moreover, $c_{\overline a b}<w_b$, so the seller gains from trade at every effort level. The buyer receives payoff $\overline v_b$ from a good output and payoff $\underline v_b$ from a bad output, where $\overline v_b > \underline v_b$. Her expected payoff from trade $b$ and seller action $a$ is therefore $v_{ab}:=p_{ab}\overline v_b+(1-p_{ab})\underline v_b$. I assume that the probability $p_{ab}$ is strictly increasing and (weakly) concave in effort $a$, and the parameters $(p_{ab}, \overline v_b, \underline v_b)$ are such that $v_{0b} < 0 < v_{\overline a b}$. Thus, a trade in which the seller shirks is not attractive to the buyer, whereas a trade in which the seller exerts enough effort is.

Let $Z:=\{0\}\cup( (B \setminus \{0\}) \times \{\overline q,\underline q\})$ be the set of public outcomes in a period, recording the buyer's action and output (or lack thereof). Let $Y:=\{0\}\cup( (B \setminus \{0\}) \times A \times\{\overline q,\underline q\})$ additionally record the seller's action. At the beginning of period $t$, the current buyer's history belongs to $Z^t$ and  the seller's history belongs to $Y^t$.

Let $u_t$ and $v_t$ denote the seller's and the buyer's realized payoff in period $t$. The seller's payoff and the ``average'' buyer's payoff are, respectively,
\begin{align*}
U:=(1-\gd)\bE\left[\sum_{t=0}^\infty\gd^t u_t\right]\!, \quad
V:=(1-\gd)\bE\left[\sum_{t=0}^\infty\gd^t v_t\right]\!,
\end{align*}
where expectations are taken under the strategy profile of interest. It is useful to interpret $\gd$ as the probability in each period that the seller continues to operate in the next period. Throughout, I refer to $(U,V)$ as a payoff vector.

I use perfect public equilibrium (PPE, \citealp{fudenberg1994folk}), in which the players' strategies depend only on the public histories, as the solution concept: each buyer's strategy specifies a distribution over her actions after each public history, whereas the seller's strategy specifies a distribution over his actions after each public history and the current buyer's trade choice. I explain at the end of this section that this public-strategy restriction is without loss of generality.

The seller's effort incentives are intrinsically dynamic: in any trade chosen by the buyer, he weighs the immediate cost of greater effort against the gain that greater effort generates by making a good output more likely. This gain depends on the difference between his continuation payoffs after good and bad outputs. \cref{sec:proofprelim} provides details on the seller's incentive constraint for each effort level. In contrast, in a one-shot interaction, he shirks after every trade that the buyer chooses, so the buyer rejects trade. Backward induction rules out trade at every finite horizon as well.

The set of feasible and individually rational payoff vectors is
\begin{align}\label{eq:calF}
\mathcal F := \operatorname{co}\left(\{(0,0)\} \cup  \bigcup_{b=1}^{\overline b}\mathcal F_b\right)
\end{align}
where, for each trade $b=1,\dots,\overline b$, 
\begin{align*}
\mathcal F_b := \operatorname{co}((u_{0b},v_{0b}),\ldots, (u_{\overline a b},v_{\overline a b}))\cap\bR_+^2
\end{align*}
is the set of feasible and individually rational payoff vectors conditional on the buyer taking trade $b$; it is a subset of $\bR_+^2$ because in each period, the seller's payoff is nonnegative and each buyer can guarantee payoff zero by rejecting trade. Therefore, the constraint on buyer participation is imposed separately for each available trade, not merely on average across trades. Let $\mathcal L$ denote the (Pareto-)efficiency frontier of $\mathcal F$, and let 
\begin{align*}
\overline U:=\max_{(U,V)\in\mathcal F}U    
\end{align*}
denote the seller's highest payoff in $\mathcal F$. It equals his mixed-action Stackelberg payoff, namely the payoff he could virtually achieve with commitment power.\footnote{Formally, let $\ga=(\ga_b)_{b=1}^{\overline b}\in\Delta(A)^{\overline b}$ specify the seller's effort distribution in each trade, and let $\mathrm{BR}(\ga)\subseteq B$ denote the set of a buyer's best responses to $\ga$. Write $u(\ga,b):=\sum_{a\in A}\ga_b(a)u_{ab}$ for $b>0$, and set $u(\ga,0):=0$. The seller's mixed-action Stackelberg payoff is
\begin{align*}
\sup_{\ga\in\Delta(A)^{\overline b}}
\min_{b\in\mathrm{BR}(\ga)}
u(\ga,b)
=\overline U.
\end{align*}
} Any payoff vector in $\mathcal F$ that gives the seller $\overline U$ gives the buyer zero. This is because for each trade $b$, increasing effort from zero to its maximum lowers the seller's payoff and raises the buyer's expected payoff from negative to positive. \cref{fig:feas} illustrates. 


\begin{figure}[t!]
\centering
\begin{subfigure}[b]{0.46\textwidth}
\centering
\begin{tikzpicture}[
x=1.05cm,
y=1.05cm,
scale=0.8,
font=\normalsize,
thick,
>=stealth'
]
\draw[fill=gray!30] (0,0) -- (4,5) -- (6,0) -- cycle;

\draw[->] (0,0) -- (0,5.5) node[left] {$V$};
\draw[->] (-0.5,0) -- (6.5,0) node[below] {$U$};

\node at (4,5.4) {$(u_{11},v_{11})=(4,5)$};
\node at (5.5,-0.3) {$\overline U=6$};
\node at (-0.2,-0.25) {$0$};

\node at (3.5,2.2) {$\mathcal F$};
\draw[ultra thick] (4,5) -- (6,0);
\node at (5,4) {$\mathcal L$};
\end{tikzpicture}
\caption{Binary actions on both sides, with $w_1=7$, $(p_{01},p_{11})=(1/4,3/4)$, $(c_{01},c_{11})=(0,3)$, and buyer payoffs $\overline v_1=35/4$ and $\underline v_1=-25/4$.\\}
\label{fig:feasiblepayoffs}
\end{subfigure}
~
\begin{subfigure}[b]{0.46\textwidth}
\centering
\begin{tikzpicture}[
x=1.05cm,
y=1.05cm,
scale=0.8,
font=\normalsize,
thick,
>=stealth'
]
\draw[fill=gray!30] (0,0) -- (1,4) -- (5,3) -- (6,0) -- cycle;

\draw[->] (0,0) -- (0,5.5) node[left] {$V$};
\draw[->] (-0.5,0) -- (6.5,0) node[below] {$U$};

\draw[ultra thick] (1,4) -- (5,3) -- (6,0);
\node at (5.8,2.5) {$\mathcal L$};

\node at (2.2,4.35) {$(u_{11},v_{11})=(1,4)$};
\node at (5.5,3.65) {$(u_{12},v_{12})=(5,3)$};
\node at (5.5,-0.3) {$\overline U=6$};
\node at (-0.2,-0.25) {$0$};
\node at (3,2) {$\mathcal F$};
\end{tikzpicture}
\caption{Three buyer actions and two seller actions, with $(w_1,w_2)=(9,7)$, $p_{0b}=1/4$ and $p_{1b}=3/4$ for $b=1,2$, $(c_{01},c_{11})=(0,8)$, $(c_{02},c_{12})=(0,2)$, and buyer's payoffs $(\overline v_1,\underline v_1)=(8,-8)$ and $(\overline v_2,\underline v_2)=(6,-6)$.}
\label{fig:feasiblepayoffs-many-panel}
\end{subfigure}
\caption{Feasible and individually rational payoffs, and the efficiency frontier\label{fig:feas}}
\end{figure}

At discount factor $\gd$, let $\mathcal E_\gd$ denote the set of PPE payoff vectors. This set is nonempty and contains $(0,0)$, sustained by a PPE that prescribes that buyers permanently reject trade and that the seller shirk after any unexpected trade.

\cref{prop:Flbenchmark} characterizes the set $\mathcal E_\gd$ as the seller becomes patient. 
For nonempty sets $K,K'\subseteq \bR^2$, let $d_H(K,K')$ denote their Hausdorff distance and write $\lim_{\gd\ra1}K_\gd=K$ if $d_H(K_\gd,K) \ra 0$ as $\gd \ra 1$.

\begin{prop*}\label{prop:Flbenchmark}
There exists $\overline\gk_{\mathrm{PPE}}>0$, independent of $\gd$, such that for every $\gd\in(0,1)$, the seller's payoff in any PPE is at most $\overline U-\overline\gk_{\mathrm{PPE}}$. Moreover,
\begin{align}\label{eq:limitPPE}
\lim_{\gd\ra1}\mathcal E_\gd
=\left\{(U,V)\in\mathcal F:
U\le\overline U-\overline\gk_{\mathrm{PPE}}\right\}
\subsetneq\mathcal F.
\end{align}
\end{prop*}


\cref{prop:Flbenchmark} extends the result of a long-lived seller's patient-limit PPE payoffs in \citet{fudenberg1994efficiency} to the patient-limit joint payoffs of the seller and the average buyer. It shows that, first, the seller's highest PPE payoff remains bounded away from his highest feasible payoff $\overline U$ by at least $\overline\gk_{\mathrm{PPE}}$, even as he becomes arbitrarily patient and, second, every interior payoff vector $(U,V)$ in $\mathcal F$ satisfying $U<\overline U-\overline\gk_{\mathrm{PPE}}$ can be attained by a PPE when the seller is sufficiently patient. The loss $\overline\gk_{\mathrm{PPE}}$, for which the proof gives an explicit formula, measures the severity of the seller's moral-hazard problem; it can equal $\overline U$, in which case the seller's moral hazard is so severe that the limiting payoff set in \eqref{eq:limitPPE} is $\{(0,0)\}$. The proof is in \cref{sec:proofs}, as are the remaining proofs unless stated otherwise.


The limiting PPE payoff set may intersect the efficiency frontier $\mathcal L$. This occurs if and only if the payoff loss $\overline\gk_{\mathrm{PPE}}$ is low enough that $\overline U-\overline\gk_{\mathrm{PPE}}$ is at least the lowest seller payoff on $\mathcal L$. \cref{fig:feasiblepayoffs-many} illustrates a setting in which this is the case, while \cref{fig:feasiblepayoffs-many-separated} illustrates one in which it is not. In the classic reputation setting where both sides have binary actions (\egg \citealp{ekmekci2011sustainable}; \citealp{levine2021reputation}), this intersection must be empty.\footnote{Indeed, $\overline U-\overline\gk_{\mathrm{PPE}}<u_{11}$, whereas every point of $\mathcal L$ gives the seller at least $u_{11}$. Their distance is therefore at least $u_{11}-(\overline U-\overline\gk_{\mathrm{PPE}})>0$, uniformly in $\gd$.} \cref{fig:higheffsetfull} illustrates. If the intersection happens to be nonempty, to the extent that only the average buyer's payoff is concerned, mediation adds little when the seller is sufficiently patient.


\begin{figure}[t!]
\centering
\begin{subfigure}[b]{0.46\textwidth}
\centering
\begin{tikzpicture}[
x=1.05cm,
y=1.05cm,
scale=0.85,
font=\normalsize,
thick,
>=stealth'
]
\path[fill=gray!30] (0,0) -- (1,4) -- (5,3) -- (6,0) -- cycle;
\path[fill=gray!70] (0,0) -- (1,4) -- (4,{13/4}) -- (4,0) -- cycle;

\draw (0,0) -- (1,4) -- (5,3) -- (6,0) -- cycle;
\draw[ultra thick] (1,4) -- (5,3) -- (6,0);
\draw[dotted] (4,0) -- (4,{13/4});

\draw[->] (0,0) -- (0,5.5) node[left] {$V$};
\draw[->] (-0.5,0) -- (6.5,0) node[below] {$U$};
\node at (-0.2,-0.25) {$0$};

\node at (2,4.35) {$(u_{11},v_{11})=(1,4)$};
\node at (5.8,3.6) {$(u_{12},v_{12})=(5,3)$};
\node at (3.5,-0.3) {$\overline U-\overline\gk_{\mathrm{PPE}}=4$};
\node at (5.7,-0.3) {$\overline U=6$};

\node at (5,1.3) {$\mathcal F$};
\node at (2.2,2) {$\displaystyle\lim_{\gd\ra1}\mathcal E_\gd$};
\node at (5.8,2) {$\mathcal L$};
\end{tikzpicture}
\caption{Intersecting the efficiency frontier, with $(w_1,w_2)=(9,7)$, $p_{0b}=1/4$ and $p_{1b}=3/4$ for $b=1,2$, $(c_{01},c_{11})=(0,8)$, and $(c_{02},c_{12})=(0,2)$. The buyer's payoffs are $(\overline v_1,\underline v_1)=(8,-8)$ and $(\overline v_2,\underline v_2)=(6,-6)$.\\}
\label{fig:feasiblepayoffs-many}
\end{subfigure}
~
\begin{subfigure}[b]{0.46\textwidth}
\centering
\begin{tikzpicture}[
x=1.05cm,
y=1.05cm,
scale=0.85,
font=\normalsize,
thick,
>=stealth'
]
\path[fill=gray!30] (0,0) -- (1,4) -- (5,3) -- (6,0) -- cycle;
\path[fill=gray!70] (0,0) -- (0.75,3) -- (0.75,0) -- cycle;

\draw (0,0) -- (1,4) -- (5,3) -- (6,0) -- cycle;
\draw[ultra thick] (1,4) -- (5,3) -- (6,0);
\draw[dotted] (0.75,0) -- (0.75,3);

\draw[->] (0,0) -- (0,5.5) node[left] {$V$};
\draw[->] (-0.5,0) -- (6.5,0) node[below] {$U$};
\node at (-0.2,-0.25) {$0$};

\node at (2,4.35) {$(u_{11},v_{11})=(1,4)$};
\node at (5.8,3.6) {$(u_{12},v_{12})=(5,3)$};
\node at (1.5,-0.3) {$\overline U-\overline\gk_{\mathrm{PPE}}=3/4$};
\node at (5.7,-0.3) {$\overline U=6$};

\node at (4,1) {$\mathcal F$};
\node at (5.8,2) {$\mathcal L$};
\node[anchor=west] at (1.6,2) {$\displaystyle\lim_{\gd\ra1}\mathcal E_\gd$};
\draw[thin,->] (1.6,1.85) -- (0.45,1.3);
\end{tikzpicture}
\caption{Separation from the efficiency frontier, with $(w_1,w_2)=(9,7)$, $p_{0b}=1/4$ and $p_{1b}=49/100$ for $b=1,2$, $(c_{01},c_{11})=(0,8)$, and $(c_{02},c_{12})=(0,2)$. The buyer's payoffs are $(\overline v_1,\underline v_1)=(21,-37/3)$ and $(\overline v_2,\underline v_2)=(63/4,-37/4)$.}
\label{fig:feasiblepayoffs-many-separated}
\end{subfigure}
\caption{Limiting PPE payoffs, with three buyer actions and two seller actions.}
\end{figure}

\begin{figure}[t!]
\centering
\begin{tikzpicture}[
x=1.05cm,
y=1.05cm,
scale=0.8,
font=\normalsize,
thick,
>=stealth'
]
\path[fill=gray!30] (0,0) -- (4,5) -- (6,0) -- cycle;
\path[fill=gray!70] (0,0) -- (2.5,{25/8}) -- (2.5,0) -- cycle;

\draw (0,0) -- (4,5) -- (6,0) -- cycle;
\draw[ultra thick] (4,5) -- (6,0);
\draw[dotted] (2.5,0) -- (2.5,{25/8});

\draw[->] (0,0) -- (0,5.5) node[left] {$V$};
\draw[->] (-0.5,0) -- (6.5,0) node[below] {$U$};
\node at (-0.2,-0.25) {$0$};

\node at (4,5.4) {$(u_{11},v_{11})=(4,5)$};
\node at (2.5,-0.4) {$\overline U-\overline\gk_{\mathrm{PPE}}=5/2$};
\node at (5.6,-0.4) {$\overline U=6$};

\node at (1.5,0.6) {$\displaystyle\lim_{\gd\ra1}\mathcal E_\gd$};
\node at (3.5,2.2) {$\mathcal F$};
\node at (5,4) {$\mathcal L$};
\end{tikzpicture}
\caption{Limiting PPE payoffs with binary actions on both sides, with $w_1=7$, $(p_{01},p_{11})=(1/4,3/4)$, $(c_{01},c_{11})=(0,3)$, and buyer payoffs $\overline v_1=35/4$ and $\underline v_1=-25/4$.}
\label{fig:higheffsetfull}
\end{figure}

In the following, I sketch the proof of \cref{prop:Flbenchmark}, proceeding in three steps. The first step establishes the seller-payoff bound in \eqref{eq:limitPPE}, which follows from standard recursive methods \citep{fudenberg1994efficiency}. Intuitively, in any continuation attaining the seller's highest PPE payoff, a current good output cannot be rewarded with an even higher continuation payoff, because every continuation is itself a PPE. Positive effort must therefore be sustained by lower continuation payoffs after bad outputs. Because outputs are noisy, these lower continuation payoffs impose a payoff loss that persists as the seller becomes arbitrarily patient.

The second step shows that, for every sufficiently high discount factor, all payoff vectors in $\{(U,0):U\in[0,\overline U-\overline\gk_{\mathrm{PPE}}]\}$ are PPE payoff vectors. If this upper bound is $0$, the claim is immediate. If it is positive instead, the payoff vector $(\overline U-\overline\gk_{\mathrm{PPE}}, 0)$ is attained by a PPE that begins with some trade $b^*$ in which the seller mixes between the lowest effort that makes the buyer willing to take trade $b^*$ and the effort immediately below it so that, given the mixture, the buyer breaks even by trading $b^*$, until the seller produces a bad output and public randomization triggers permanent rejection with a probability chosen to make the seller indifferent between the efforts in his mixture. For any payoff vector in $\{(U,0):U\in[0,\overline U-\overline\gk_{\mathrm{PPE}}]\}$, a suitable public randomization between this PPE and another PPE attaining $(0,0)$ gives a PPE attaining  that vector.

The third step fixes $(U,V)$ in the interior of $\mathcal F$ with $U<\overline U-\overline\gk_{\mathrm{PPE}}$ and constructs a PPE attaining it for sufficiently high $\gd$. Choose $\widetilde V>V$ with $(U,\widetilde V)$ in the interior of $\mathcal F$. I first construct a PPE attaining seller payoff $U$ and an average-buyer payoff that approaches $\widetilde V$ as the seller becomes sufficiently patient. This PPE has two phases, beginning with a main phase. During this phase, public randomization selects buyer actions and possibly mixed seller effort strategies that make the buyer willing to follow her prescribed action and yield expected current payoffs $(U,\widetilde V)$. Starting from initial seller continuation payoff $U$, the continuation payoffs after good and bad outputs are chosen to satisfy the corresponding incentive constraint and the promise-keeping condition at each history. The phase continues while the seller's continuation payoff remains in a fixed interval around the seller's initial payoff. Upon exiting the interval with continuation payoff $U'$, play enters a residual phase, switching to a PPE attaining $(U',0)$, which exists by the second step. As the seller becomes patient, the required changes in the seller's continuation payoff at the end of each period in the main phase vanish because one-shot deviation gains become negligible, and the expected discounted weight of the residual phase tends to zero. The average buyer's payoff therefore converges to $\widetilde V>V$. For sufficiently high $\gd$, public randomization between this PPE attaining seller payoff $U$ and average-buyer payoff near $\widetilde V$ and the PPE attaining $(U,0)$ gives a PPE attaining $(U,V)$. This completes the argument for \cref{prop:Flbenchmark}.

\cref{prop:Flbenchmark} would continue to hold if the solution concept were sequential equilibrium in which the seller's strategy can additionally depend on his past private actions. This is because monitoring has a product structure and buyers' actions are publicly observable \citep{fudenberg1994efficiency}, so that the PPE and sequential-equilibrium payoff sets coincide. The known advantages of private strategies over public ones therefore do not apply in my model.

My assumption that buyers observe the full public history gives the most demanding case for mediation to expand equilibrium payoffs, in the following sense. If instead each buyer observed only the $K$ most recent public outcomes, for some $K\ge0$, the equilibrium payoff set without mediation would be a subset of $\mathcal E_\gd$ for each discount factor $\gd$. The reason parallels \citet{kandori1992use}.\footnote{\citet{kandori1992use} shows that the set of PPE payoffs is weakly increasing in the informativeness of the public signal about current actions; here the comparison concerns information about the past.} Fix a sequential equilibrium in this setting. At each date, buyers condition their strategies on these recent outcomes (including the public randomization draws in those most recent past periods, as well as the current draw). The seller may condition on more. Consider, nonetheless, replacing his effort strategy in each possible trade with the distribution that the buyer expects in the original equilibrium conditional on her information. This remains a best reply for the seller: given buyers' strategies, histories at the same date with the same recent outcomes present the same continuation problem, since no future buyer observes the older outcomes. The replacement preserves every buyer's expected payoff from each action and, by induction, the distribution of what buyers observe. The resulting strategy profile therefore attains the same ex ante payoffs. When buyers observe the full public history, these strategies remain optimal while ignoring older outcomes and thus constitute a PPE with the same payoffs.
\section{Mediated efficiency}
\label{sec:model}

I now introduce mediation into the benchmark model. Following \citet{forges1986approach} and \citet{myerson1986multistage}, a mediator commits ex ante to a communication device that, before the buyer moves in each period $t$, sends a private message $m^1_t\in M^1$ to the seller and a private message $m^2_t\in M^2$ to the buyer; the mediator freely chooses finite message sets $M^1$ and $M^2$.\footnote{Finite message sets suffice for my results: my construction for \cref{thm:payoffset} uses finitely many messages, while the revelation principle allows me to restrict messages to finitely many action recommendations in \cref{prop:uniformseparation}. The results also hold with countable message sets. This extension matters for the robustness claim in the Introduction: if buyers observe only part of the public history, countable message sets allow the mediator to disclose the full history of buyer actions and outputs alongside its original messages and replicate the same equilibrium outcomes that I construct, so \cref{thm:payoffset} continues to hold.} Let $M:=M^1\times M^2$ denote the set of message profiles. Denote the set of mediator histories by $H_t:=(M\times Z)^t$, with $h_0:=\varnothing$; a mediator history records past messages, buyer actions, and outputs, but not seller actions.\footnote{Even if the mediator does not observe buyer actions and outputs directly, it can freely elicit this information from the short-lived buyers, who are willing to report it truthfully in equilibrium. Note also that the mediator does not elicit information from the seller about his actions. Allowing for this possibility does not affect my results: the seller cannot benefit from reporting that he has deviated from the action prescribed by the mediator.} For consistency, the public randomization device remains available, but it plays no role. Buyer actions and outputs remain publicly observable.

\begin{def*}\label{def:commdevice}
A communication device is a collection $D\equiv(D_t)_{t=0}^\infty$, where
\begin{align*}
D_t:H_t\ra\Delta(M)    
\end{align*}
specifies a lottery over current message profiles at each mediator history.
\end{def*}

I say that a game (between the seller and the buyers) is induced by communication device $D$ if the mediator chooses $D$. The device is known to all players. The mediator cannot compel trade or enforce positive effort. Every buyer remains free to reject trade, and the seller remains free to choose his effort.

In the game induced by any communication device,  the seller's period-$t$ private history belongs to $(M^1\times Y)^t$. His strategy specifies a distribution over $A$ for each such history, current message, and trade $b>0$ chosen by the current buyer. The current buyer observes the public history in $Z^t$ and her message; her strategy specifies a distribution over $B$ given this history and the message.

With mediation, the standard solution concept is communication equilibrium: a pair $(D,\gs)$ is a \emph{communication equilibrium} (CE) if $\gs$ is a Nash equilibrium of the game induced by $D$. Communication equilibrium is, however, generally too permissive in a dynamic setting, since Nash equilibrium does not require sequential rationality. The equilibria I construct are strong sequential communication equilibria, in the sense of \citet{gerardi2007sequential}, and hence satisfy sequential rationality.

\begin{def*}\label{def:strongcommeqm}
A pair $(D,\gs)$ is a \emph{strong sequential communication equilibrium} (SSCE) if there exists a belief system $\mu$ over histories of play satisfying:
\begin{enumerate}\itemsep0em
\item \emph{Sequential rationality}: given $\mu$, the strategies in $\gs$ are the seller's and the buyers' mutual best responses at every private history and current message and, for the seller, after every trade chosen by the buyer.
\item \emph{Consistency}: there exists a sequence of completely mixed strategy profiles $(\gs^n)_{n=1}^\infty$ converging pointwise to $\gs$ such that the sequence of belief systems induced by $D$ and $(\gs^n)_{n=1}^\infty$ converges pointwise to $\mu$.\footnote{\citet{gerardi2007sequential} define SSCE in finite Bayesian games. Here I retain their requirement that consistency be obtained by perturbing the players' strategies while holding the communication device fixed. By contrast, sequential communication equilibrium (SCE, \citealp{myerson1986multistage}) also permits mediator trembles. SSCE is a refinement of SCE.}
\end{enumerate}
\end{def*}


My main result, \cref{thm:payoffset}, establishes that every interior feasible and individually rational payoff vector is an SSCE payoff vector when the seller is sufficiently patient. In particular, mediation can bring the seller's payoff arbitrarily close to any of his payoffs on the efficiency frontier $\mathcal L$, including his mixed-action Stackelberg payoff $\overline U$.  Let $\mathcal S_\gd\subseteq\mathcal F$ be the set of SSCE payoff vectors at discount factor $\gd$.

\begin{thm*}\label{thm:payoffset}
For every $(U,V)$ in the interior of $\mathcal F$, there exists $\gd_0<1$ such that $(U,V)\in\mathcal S_\gd$ for every $\gd\in(\gd_0,1)$. Therefore
\begin{align*}
    \lim_{\gd \ra 1} \mathcal S_\gd = \mathcal F.
\end{align*}
\end{thm*}

Private communication lets the mediator induce continuations prescribing less effort while buyers remain willing to trade. Buyers expect enough effort across hidden histories, although they would reject at some of these histories if they knew which one had occurred. These low-effort trades give the seller higher current payoffs and provide additional rewards for good outputs, overcoming the constraint on his equilibrium payoff in \cref{prop:Flbenchmark}. \cref{thm:payoffset} shows that these rewards are powerful enough to help attain every interior payoff vector in $\mathcal F$ when the seller is sufficiently patient.

In the rest of this section, I sketch the proof of \cref{thm:payoffset}. For each interior payoff vector in $\mathcal F$, I construct an SSCE that attains it when the seller is sufficiently patient. For each trade $b>0$, let $\mathcal L_b$ denote the (Pareto-)efficiency frontier of $\mathcal F_b$. I first show that SSCE payoffs can approach every point on $\mathcal L_b$ as the seller becomes patient. Consider a target payoff vector $(U^*,V^*)$ on $\mathcal L_b$ with $V^*>0$, generated by randomizing over adjacent efforts $a^*-1$ and $a^*$, with probability $\gl\in(0,1)$ assigned to $a^*$ and complementary probability assigned to $a^*-1$ in trade $b$:
\begin{align}\label{eq:UVstar}
(U^*,V^*) = \gl(u_{a^*b},v_{a^*b}) + (1-\gl)(u_{(a^*-1)b},v_{(a^*-1)b}).
\end{align}
Let $\pi:=\gl p_{a^*b} + (1-\gl)p_{(a^*-1)b}$ denote the good-output probability induced by this mixture. The buyer's payoff $V^*$ can therefore be written as $V^*=\pi\overline v_b+(1-\pi)\underline v_b$. I write the target as $(U^*_{b,\pi},V^*_{b,\pi})$ to indicate its dependence on $\pi$. Once SSCE payoffs are shown to approach this target as the seller becomes patient, it follows that SSCE payoffs approach every point on $\mathcal L_b$. This is because, as $a^*$ and $\gl$, and therefore $\pi$, vary, the corresponding targets form a dense subset of $\mathcal L_b$.

Fix a generic, sufficiently small $\ve\in(0,U^*_{b,\pi})$. I construct an SSCE attaining a payoff vector that approaches the target $(U^*_{b,\pi},V^*_{b,\pi})$ as $\ve$ vanishes, where, for each $\ve$, the seller is sufficiently patient. In the SSCE, the communication device sends action recommendations, potentially alongside auxiliary messages; the action recommendation to the seller is contingent on the buyer's action. The strategy profile is obedient, in the sense that each player follows these recommendations at every private history.

The communication device uses a finite ladder of trade states and one absorbing no-trade state. Trade states are indexed by $n=1,\dots,N$ from bottom to top, where $N \ge 3$ is determined below. Initially, the state is drawn from a distribution $\rho$ over the trade states. At trade state $n$, the device tells the seller, but not the buyer, the state. It also (privately) recommends that the buyer choose $b$, and that the seller exert effort $a_n \in A$ in trade $b$ or shirk in any other trade. The prescribed efforts $(a_1,\dots,a_N)$, chosen below, satisfy $a_1=\overline a\ge a_2\ge\cdots\ge a_N$, $p_{a_Nb}<\pi$, and $p_{\overline ab}>\pi$. Therefore, higher trade states prescribe (weakly) lower efforts, with the lowest prescribed effort being sufficiently low and the highest one being the maximal one. At the no-trade state, the device recommends rejection and shirking in any unexpected trade. An observable deviation, which only a buyer can carry out, sends the next state to the no-trade state. I choose the initial distribution $\rho$, the state transitions, and the prescribed efforts $(a_1,\dots,a_N)$ so that in the resulting SSCE, the seller's continuation payoff at each trade state converges to $U^*_{b,\pi}-\ve$ as $\gd\ra1$.


I first describe the basic structure of the state transitions, before specifying the parameters in detail. After each prescribed trade, the device follows a local transition rule, described momentarily, with probability $1-\xi$ and privately redraws the next trade state from distribution $\rho$ with probability $\xi$, where $\xi$ will be chosen to be positive but small. An exception is a bad output at the bottom state, after which the device enters the no-trade state with some probability $\eta\in(0,\xi)$ that I choose below, redraws from distribution $\rho$ with probability $\xi-\eta$, and follows the local rule with probability $1-\xi$. Under the local rule, an interior state moves one rung up the ladder with probability $1-\pi$ after a good output and one rung down with probability $\pi$ after a bad output. At the top state, a bad output moves the state down with probability $(\pi-p_{a_Nb})/(1-p_{a_Nb})$. At the bottom state, a good output moves the state up with probability $(p_{\overline ab}-\pi)/p_{\overline ab}$. The two probabilities are well-defined, as the prescribed efforts will be chosen to satisfy $\pi > p_{a_Nb}$ and $p_{\overline ab} > \pi$. In other cases, the local rule leaves the state unchanged. 

I choose the initial distribution $\rho$ so that, under obedience, if the current trade state has distribution $\rho$, then the next state also has distribution $\rho$ conditional on either output and continued trade, given the state transitions and the prescribed effort profile. Absent rejection, such a distribution exists under the local transition rule and redraws. Rejection does not disturb this property because it takes its probability from the redraw: after a bad output at the bottom state, the rejection probability $\eta$ that would otherwise have redrawn the next trade state from $\rho$ sends the device to the no-trade state instead. The probability removed from the trade states is therefore proportional to $\rho$, so their distribution conditional on continued trade is still $\rho$.

The choice of $\rho$ ensures the buyers' obedience. By construction, for each interior trade state $n$, the expected change in the state index is $p_{a_nb}(1-\pi)-(1-p_{a_nb})\pi = p_{a_nb}-\pi$ under the local transition rule; for the top and bottom trade states $n$, an analogous calculation gives the same expected change $p_{a_nb}-\pi$ under the local rule. Since the distribution $\rho$ is stationary under the local rule, the expected change in the state index averaged under $\rho$ is zero: $\sum_{n=1}^N \rho_n ( p_{a_nb} - \pi) = 0$. Because each buyer recommended trade $b$ has posterior belief $\rho$ over the trade states after every public history, she believes that the output will be good with probability $\pi$:
\begin{align}\label{eq:buyerbelief}
\sum_{n=1}^N \rho_n p_{a_nb}  = \pi.
\end{align}
Consequently, her expected payoff from trade $b$ is $V^*_{b,\pi}$. Because $V^*_{b,\pi}>0$, she strictly prefers to obey a recommendation to trade $b$ over any deviation, which would result in either no trade or the seller's shirking in an unexpected trade. In contrast, a buyer recommended rejection expects shirking in every trade and strictly prefers rejection.

I choose the prescribed efforts from the top down the ladder. I first introduce the necessary notation. Let $c_b^*:=(1-\gl)c_{(a^*-1)b}+\gl c_{a^*b}$ denote the expected effort cost under the target mixture, namely effort $a^*$ with probability $\gl$ and effort $a^*-1$ otherwise. For each $a >0$, let
\begin{align}\label{eq:Hab}
J_{ab}:=c_{ab}+(1-p_{ab})
\frac{c_{ab}-c_{(a-1)b}}{p_{ab}-p_{(a-1)b}}
\end{align}
and set $J_{0b}:=0$. Because effort cost is convex and good-output probability is concave in effort, $J_{ab}$ is nondecreasing in $a$. The quantity $J_{ab}$ in \eqref{eq:Hab} combines the cost of effort $a$ with the minimum expected payoff loss after a bad output needed to sustain that effort, measured in per-period payoff units. To understand this payoff loss, consider a state $n$ in which the seller is prescribed effort $a>0$  and let $U_{n,\overline q}$ and $U_{n,\underline q}$ denote his continuation payoffs after good and bad outputs. The seller must not have a profitable deviation to effort $a-1$, so that the following incentive constraint holds, which ensures that the cost of the additional unit of effort is sufficiently small relative to the good-minus-bad continuation-payoff difference:
\begin{align}\label{eq:aic}
	\frac{c_{ab}-c_{(a-1)b}}{p_{ab}-p_{(a-1)b}} \le \frac{\gd}{1-\gd}(U_{n,\overline q}-U_{n,\underline q}).
\end{align}
The left side---the effort-cost saving per unit fall in the good-output probability from such a deviation---is therefore the minimal scaled good-minus-bad continuation-payoff difference that makes a deviation to $a-1$ unprofitable, where the scaling by $\gd/(1-\gd)$ expresses this difference in per-period payoff units so that the two sides of \eqref{eq:aic} are comparable.\footnote{This scaling also matters for the limit: as $\gd\ra1$, the unscaled differences between the seller's continuation payoffs at adjacent trade states vanish, whereas the scaled differences converge to finite, nonzero limits. This property is used in defining the gaps $g_n$ below.} Multiplying the left side of \eqref{eq:aic} by the bad-output probability $1-p_{ab}$ under effort $a$ then gives the second term in $J_{ab}$, capturing the minimal expected continuation payoff loss the seller incurs, in per-period payoff units, that keeps him indifferent between effort $a$ and deviating to effort $a-1$. Deterring every deviation from effort $a$ requires, in particular, deterring the deviation to $a-1$, so this is the minimal loss. Finally, for each $n<N$, let $g_n$ denote the joint limit, as $(\xi, \gd) \ra (0,1)$, of the gap between the seller's continuation payoffs at states $n+1$ and $n$ multiplied by $\gd/(1-\gd)$, so that the gap is expressed in per-period payoff units. Set $g_N :=0$. For each state $n$, from $N$ through 2, the prescribed effort $a_n$ is
\begin{align}\label{eq:selection}
a_n:=\max \{a\in A: w_b + (1-\pi)g_n -J_{ab} >U^*_{b,\pi}-\ve \}.
\end{align}
Therefore, it is the highest effort for which the seller's current trade payoff, plus the good-output continuation reward and minus the minimum expected bad-output penalty needed to deter a deviation, strictly exceeds the target state-$n$ continuation payoff $U^*_{b,\pi}-\ve$ in the joint limit. Because $g_N=0$, $a_N$ is readily determined by \eqref{eq:selection}. The other prescribed efforts are determined recursively together with the gaps. Specifically, in the joint limit, the seller's required continuation payoff $U^*_{b,\pi}-\ve$ at state $N$ satisfies the promise-keeping condition
\begin{align*}
U^*_{b,\pi}-\ve = w_b-c_{a_Nb}-(\pi-p_{a_Nb})g_{N-1}.
\end{align*}
The right side of this equation follows because at this state, the seller's current trade payoff is $w_b-c_{a_Nb}$ and, in the joint limit, the unconditional probability of moving down the ladder is $\pi-p_{a_Nb}$ and this move costs the seller $g_{N-1}$ in per-period payoff units. This promise-keeping condition pins down the value of $g_{N-1}$ and, in turn, the value of $a_{N-1}$ by \eqref{eq:selection}. Recursively, for each $n=N-1,\dots,2$, once $a_n$ is determined, the seller's promise-keeping condition at state $n$ determines $g_{n-1}$, which then determines $a_{n-1}$ unless the next state is the bottom one, where I set $a_1=\overline a$.

The recursion delivers the required properties for the prescribed effort choices: $a_1=\overline a\ge a_2\ge\cdots\ge a_N$, $p_{a_Nb}<\pi$, and $p_{\overline ab}>\pi$. To see the first chain of inequalities, note that the gap $g_n$ is strictly decreasing in $n$. In each interior state $n$, consider a one-shot deviation in which the seller plays the target mixture, choosing effort $a^*$ with probability $\gl$ and effort $a^*-1$ otherwise, and then follows the prescribed strategy thereafter. This gives him expected current payoff $U^*_{b,\pi}$, which exceeds his limiting continuation payoff at that state by $\ve$. In the joint limit, the local transition rule makes upward and downward moves equally likely under this deviation, each with probability $\pi(1-\pi)$. To prevent this deviation, the fall in continuation payoff from moving down one state must therefore be strictly larger than the rise in continuation payoff from moving up one state, leading to a larger gap at a lower state. Consequently, for each $n$, every effort satisfying $J_{ab}<c_b^*+\ve+(1-\pi)g_n$ also satisfies $J_{ab}<c_b^*+\ve+(1-\pi)g_{n-1}$. Thus, by \eqref{eq:selection}, the prescribed effort at a lower state cannot be lower.  Next, $p_{a_Nb}<\pi$. By \eqref{eq:selection} with $g_N=0$ and $J_{a_Nb}\ge c_{a_Nb}$, the prescribed effort satisfies $w_b-c_{a_Nb}>U^*_{b,\pi}-\ve$, that is, $c_{a_Nb}<c_b^*+\ve$. For sufficiently small $\ve$, the right side is below $c_{a^*b}$, so $a_N\le a^*-1$ and hence $p_{a_Nb}\le p_{(a^*-1)b}<\pi$ by the definition of $\pi$. Finally, $p_{\overline a b}>\pi$: since $\overline a\ge a^*$ and the good-output probability increases with effort, $p_{\overline a b}\ge p_{a^*b}$, and the target mixture assigns positive probability $1-\gl$ to the lower effort $a^*-1$, so $\pi<p_{a^*b}\le p_{\overline a b}$.

The ladder length $N$ is chosen such that in the above recursion, the gap $g_1$ satisfies
\begin{align}\label{eq:stoppingc}
w_b + (p_{\overline a b}-\pi)g_1 -J_{\overline a b}
>U^*_{b,\pi}-\ve.
\end{align}
Such an $N$ exists because $p_{\overline a b}>\pi$ and the gap increases by at least a fixed positive amount at each step, so \eqref{eq:stoppingc} holds after finitely many steps.

I pick the redraw probability $\xi$ to be sufficiently small that a sufficiently patient seller has no profitable deviation from the prescribed efforts. In each trade state $n$ except the bottom one, the convexity of effort cost and the concavity of good-output probability in effort imply that it suffices to rule out profitable deviations to adjacent efforts. If $a_n>0$, consider first a downward deviation. By definition of $J_{a_nb}$, if the seller were indifferent between $a_n$ and $a_n-1$, his continuation payoff in state $n$ in the joint limit as $(\xi,\gd)\ra(0,1)$ would be $w_b+(1-\pi)g_n-J_{a_nb}$. By \eqref{eq:selection}, this payoff strictly exceeds $U^*_{b,\pi}-\ve$, so the equilibrium punishment after a bad output must be stronger, which makes the seller strictly prefer $a_n$ to $a_n-1$. Similarly, if $a_n<\overline a$, consider an upward deviation. Equation \eqref{eq:selection} gives $w_b+(1-\pi)g_n-J_{(a_n+1)b}\le U^*_{b,\pi}-\ve$, with strict inequality for generic $\ve$. If the seller were indifferent between $a_n$ and $a_n+1$, his continuation payoff in state $n$ in the joint limit would therefore be strictly lower than $U^*_{b,\pi}-\ve$, so the equilibrium punishment after a bad output must be less severe, which makes the seller strictly prefer $a_n$ to $a_n+1$. At the bottom state, \eqref{eq:stoppingc} ensures that the reward from moving up after a good output, together with the rejection punishment, makes effort $\overline a$ strictly optimal. By picking $\xi$ to be sufficiently small, these strict incentives continue to hold provided that the seller is sufficiently patient. 

I next pick the rejection probability $\eta$. Since the distribution $\rho$ is stationary under the local transition rule, the required patient-limit continuation payoff $U^*_{b,\pi}-\ve$ at each trade state equals the expected current trade payoff $\sum_n\rho_nu_{a_nb}$ under $\rho$ minus the required payoff loss, measured in per-period payoff units, from the possibility of rejection. Because this latter loss is positive, $\sum_n\rho_nu_{a_nb}>U^*_{b,\pi}-\ve$. Consequently, I pick $\eta>0$ so that the seller's initial payoff is exactly $U^*_{b,\pi}-\ve$ when the seller is sufficiently patient. Because the no-trade state is absorbing, for the seller's initial payoff to remain equal to $U^*_{b,\pi}-\ve$ as $\gd \ra 1$, the probability $\eta$ must vanish as $\gd \ra 1$, so $\eta<\xi$ holds when the seller is sufficiently patient, as required. 

Once $\eta$ is picked, the specification of the communication device is complete. Given this device and the obedient strategy profile, as $\gd\ra1$, the seller's continuation payoff in any trade state indeed approaches $U^*_{b,\pi}-\ve$, as desired. This is because a redraw or rejection occurs after $1/\xi$ periods on average, payoffs earned before it occurs have vanishing weight for the seller as $\gd\ra1$, and the probability of rejection before a redraw tends to zero because $\eta\ra0$. Consequently, as $\gd\ra1$, the seller's continuation payoff in any trade state converges to his continuation payoff after a redraw from $\rho$, or equivalently, his initial payoff, which is $U^*_{b,\pi}-\ve$ by choice of $\eta$.

The obedience incentives, together with the consistency of beliefs verified in the proof, ensure that the construction yields an SSCE. It remains to verify that in this SSCE, the seller's and the average buyer's payoffs approach $(U^*_{b,\pi}, V^*_{b,\pi})$ as $\ve\to0$, with the seller being sufficiently patient for each $\ve$, and therefore SSCE payoffs can approach every point on $\mathcal L_b$ as the seller becomes patient.\footnote{At each fixed discount factor for which the construction is feasible, trade eventually ends with probability one. Under obedience, conditional on continued trade, the probability of entering the absorbing no-trade state next period is the same positive number $\rho_1(1-p_{\overline a b})\eta$ after every public history. Nevertheless, the expected normalized discounted frequency of no trade is at most $\ve/U^*_{b,\pi}$, which vanishes as $\ve\to0$.} To prove this, I first show that the expected current trade payoff $\sum_n\rho_nu_{a_nb}$ is at most $U^*_{b,\pi}$. By \eqref{eq:buyerbelief}, the prescribed efforts under $\rho$ produce the same good-output probability $\pi$ as the target mixture. By the convexity of effort cost and the concavity of the good-output probability in effort, the target mixture minimizes expected effort cost at probability $\pi$. Hence $\sum_{n=1}^N\rho_nu_{a_nb}
=w_b-\sum_{n=1}^N\rho_nc_{a_nb}
\le w_b-c_b^*
=U^*_{b,\pi}$. This upper bound implies that in this SSCE, the expected normalized discounted frequency of trade is at least $1-\ve/U^*_{b,\pi}$. Since each trade gives the buyer expected payoff $V^*_{b,\pi}$, the average buyer's payoff is at least $(1-\ve/U^*_{b,\pi})V^*_{b,\pi}$ and at most $V^*_{b,\pi}$. Because the seller's payoff is equal to $U^*_{b,\pi}-\ve$, both the seller's and average buyer's payoffs approach the target as $\ve$ vanishes.

Finally, I complete the argument for \cref{thm:payoffset}. Because every vertex of $\mathcal F$ is either the origin $(0,0)$ or a vertex on $\mathcal L_b$ for some trade $b>0$, repeating the above construction for each trade $b$ approximates every vertex of $\mathcal F$ other than $(0,0)$. The origin is attained by an SSCE that prescribes permanent rejection with shirking in unexpected trades. Consequently, for any fixed payoff vector in the interior of $\mathcal F$, the mediator can publicly randomize over the corresponding SSCEs at the outset to attain that payoff vector when the seller is sufficiently patient.

\section{The limits of recursive mediation}
\label{sec:recmain}

The SSCE construction in \cref{thm:payoffset} is nonrecursive: a continuation following a given mediator history need not itself be an SSCE, because buyers' obedience depends on their beliefs over hidden mediator histories. Indeed, for each trade $b>0$, the construction gives the buyer an expected payoff $v_{a_Nb}<0$ in the top state when the target $U^*_{b,\pi}$ is sufficiently close to the seller's highest payoff on $\mathcal L_b$ and $\ve$ is sufficiently small. If an arriving buyer knew that the previous period's trade occurred at the top state and produced a good output, her expected payoff from following a recommendation to trade $b$ would be $(1-\xi)v_{a_Nb}+\xi V^*_{b,\pi}$, because the state would not be redrawn and therefore stay at the top with probability $1-\xi$. My construction uses a sufficiently small $\xi$ that this expected payoff is negative, so the buyer would then reject trade. The continuation following this mediator history is therefore not an SSCE on its own.

In this section, I show that the nonrecursive structure in my construction is crucial: recursive mediation entails a payoff loss that persists in the patient limit. I study perfect communication equilibrium (PCE, \citealp{tomala2009perfect}), which has a recursive structure. Let $D|_{h_t}$ denote the continuation of $D$ at mediator history $h_t$: formally, $D_k|_{h_t}(h_k)=D_{t+k}(h_t h_k)$, where $h_t h_k$ concatenates the two histories. Following \citet{tomala2009perfect}, I invoke the revelation principle (\citealp{forges1986approach}; \citealp{myerson1986multistage}) to focus without loss of generality on communication devices whose messages are action recommendations and also focus on obedient strategies; again, the mediator's recommendation to the seller is contingent on the buyer's action. Let $\overline\gs$ denote an obedient strategy profile.

\begin{def*}\label{def:perfectcommeqm}
A pair $(D,\overline\gs)$ is a \emph{perfect communication equilibrium} (PCE) if $(D|_{h_t},\overline\gs)$ is a CE for every period $t$ and mediator history $h_t$.
\end{def*}

By definition, unlike SSCE, PCE has a recursive structure: every continuation of a PCE is a PCE. This recursive structure implies that PCE has a belief-free property in the spirit of \citet{ely2005belief}. Although the seller and the buyers do not observe the mediator's history, PCE requires obedience to constitute a Nash equilibrium after every such history; their obedience must therefore remain optimal under any belief over these histories consistent with their information. This does not make strategies public, however, since current recommendations remain private.

Let $\mathcal P_\gd$ denote the set of PCE payoff vectors at discount factor $\gd$. 

\begin{prop*}\label{prop:uniformseparation}
There exists $\overline\gk_{\mathrm{PCE}} \in (0, \overline \gk_{\mathrm{PPE}}]$, where $\overline \gk_{\mathrm{PPE}}$ is identified in \cref{prop:Flbenchmark}, such that for every $\gd\in(0,1)$, the seller's payoff in any PCE is at most $\overline U-\overline\gk_{\mathrm{PCE}}$. Moreover, 
\begin{align}\label{eq:limitPCE}
\lim_{\gd\ra1}\mathcal P_\gd= 
\left\{(U,V)\in\mathcal F: U \le 
\overline U - \overline \gk_{\mathrm{PCE}}
\right\} \subsetneq\mathcal F.
\end{align}
\end{prop*}

Thus, the limiting PCE payoff set consists of payoff vectors in $\mathcal F$ with seller payoff at most $\overline U-\overline\gk_{\mathrm{PCE}}$, which is bounded away from his feasible maximum $\overline U$. I provide an explicit formula for the loss $\overline\gk_{\mathrm{PCE}}$ in the proof. As with PPE in \cref{prop:Flbenchmark}, the payoff vectors on the efficiency frontier that PCE can approach in the patient limit are precisely those satisfying this bound, and there may be no such vectors.

Comparing \cref{thm:payoffset} with \cref{prop:uniformseparation} shows that the gains from nonrecursive over recursive mediation persist even as the seller becomes arbitrarily patient. This is because SSCE allows the mediator to reward the seller with trades that buyers would reject if the mediator's history were revealed, as described above, and PCE rules out such continuations.

The proof of \cref{prop:uniformseparation} is analogous to that of \cref{prop:Flbenchmark} because PCE is the mediated analogue of PPE. The main difference is that the mediator can induce a mixture over seller actions without requiring the seller to be indifferent among them, although the mixture must still make the buyer willing to follow her trade recommendation. This flexibility can raise the seller's payoff upper bound and expand the equilibrium payoff set relative to PPE, and therefore $\overline\gk_{\mathrm{PCE}} \le \overline\gk_{\mathrm{PPE}}$. Indeed, at every discount factor, every PPE outcome can be implemented by a PCE, so the PPE payoff set is contained in the PCE payoff set \citep{tomala2009perfect}. The two limiting payoff sets may nevertheless be identical. This is the case when the seller's payoff upper bound in \eqref{eq:limitPCE} is attained by some trade and degenerate mixture over efforts in that trade, so the benefit of bypassing the indifference constraint is moot; \cref{fig:ppe-pce-common-limit} illustrates. Otherwise, provided this bound is positive, the limiting PCE payoff set is a strict superset of the limiting PPE payoff set; \cref{fig:ppe-pce-strict-limit} illustrates, in which the limiting PCE payoff set intersects part of the efficiency frontier but the limiting PPE payoff set does not. \cref{fig:ppe-pce-strict-separated} gives an example in which neither limiting payoff set intersects the efficiency frontier.

\begin{figure}[t!]
\centering
\begin{subfigure}[b]{0.46\textwidth}
\centering
\begin{tikzpicture}[
x=1.65cm,
y=9cm,
scale=0.85,
font=\normalsize,
thick,
>=stealth',
dot/.style={
draw,
fill=white,
circle,
inner sep=0pt,
minimum size=4pt
}
]
\path[fill=gray!30] (0,0) -- (1,0.5) -- (4,0) -- cycle;

\path[fill=gray!70] (0,0) -- (1,0.5) -- (2,{1/3}) -- (2,0) -- cycle;

\draw (0,0) -- (1,0.5) -- (4,0) -- cycle;
\draw[ultra thick] (1,0.5) -- (4,0);

\draw[dotted] (2,0) -- (2,{1/3});
\draw[dotted] (0,{1/3}) -- (2,{1/3});
\draw[dotted] (0,0.5) -- (1,0.5);

\draw[->] (0,0) -- (0,0.64) node[left] {$V$};
\draw[->] (-0.25,0) -- (4.6,0) node[below] {$U$};
\node[below left] at (0,0) {$0$};
\node[left] at (0,0.5) {$\frac12$};
\node[left] at (0,{1/3}) {$\frac13$};

\node[dot] at (1,0.5) {};
\node at (1.5,0.565) {$(u_{2b},v_{2b})=(1,1/2)$};
\node at (2,-0.04) {$2$};
\node at (4,-0.04) {$\overline U=4$};

\node at (2.5,0.1) {$\mathcal F$};
\node at (3.55,0.16) {$\mathcal L$};
\node at (3,0.435) {$\displaystyle\lim_{\gd\ra1}\mathcal E_\gd
=\lim_{\gd\ra1}\mathcal P_\gd$};
\draw[thin,->] (2.25,0.39) -- (1.55,0.22);
\end{tikzpicture}
\caption{No expansion, with $w_b=5$, $(p_{0b},p_{1b},p_{2b})=(1/4,1/2,3/4)$, $(c_{0b},c_{1b},c_{2b})=(0,1,4)$, and buyer payoffs $\overline v_b=1$ and $\underline v_b=-1$. }
\label{fig:ppe-pce-common-limit}
\end{subfigure}
~
\begin{subfigure}[b]{0.46\textwidth}
\centering
\begin{tikzpicture}[
x=1.65cm,
y=9cm,
scale=0.85,
font=\normalsize,
thick,
>=stealth',
dot/.style={
draw,
fill=white,
circle,
inner sep=0pt,
minimum size=4pt
}
]

\path[fill=gray!30] (0,0) -- (2.5,0.5) -- (4,0) -- cycle;

\path[fill=gray!70] (0,0) -- (2,{2/5}) -- (2,0) -- cycle;

\path[fill=blue, opacity=0.2] (0,0) -- (2.5,0.5) -- (3,{1/3}) -- (3,0) -- cycle;

\draw (0,0) -- (2.5,0.5) -- (4,0) -- cycle;
\draw[ultra thick] (2.5,0.5) -- (4,0);

\draw[dotted] (2,0) -- (2,{2/5});
\draw[dotted] (3,0) -- (3,{1/3});
\draw[dotted] (0,0.5) -- (2.5,0.5);

\draw[->] (0,0) -- (0,0.64) node[left] {$V$};
\draw[->] (-0.25,0) -- (4.6,0) node[below] {$U$};
\node[below left] at (0,0) {$0$};
\node[left] at (0,0.5) {$\frac12$};
\draw (2,0) -- (2,-0.008);
\draw (3,0) -- (3,-0.008);
\node[below] at (2,-0.008) {$2$};
\node[below] at (3,-0.008) {$3$};

\node[dot] at (2.5,0.5) {};
\node at (2.5,0.565) {$(u_{2b},v_{2b})=(5/2,1/2)$};
\node at (4,-0.04) {$\overline U=4$};

\node at (3.4,0.055) {$\mathcal F$};
\node at (3.65,0.19) {$\mathcal L$};
\node at (0.9,0.35) {$\displaystyle\lim_{\gd\ra1}\mathcal E_\gd$};
\draw[thin,->] (0.95,0.295) -- (1.35,0.14);
\node[sgblue] at (3.85,0.435) {$\displaystyle\lim_{\gd\ra1}\mathcal P_\gd$};
\draw[sgblue, thin,->] (3.5,0.385) -- (2.6,0.18);
\end{tikzpicture}
\caption{Strict expansion, with $w_b=5$, $(p_{0b},p_{1b},p_{2b})=(1/4,1/2,3/4)$, $(c_{0b},c_{1b},c_{2b})=(0,1,5/2)$, and buyer payoffs $\overline v_b=1$ and $\underline v_b=-1$.}
\label{fig:ppe-pce-strict-limit}
\end{subfigure}
\caption{PCE versus PPE payoffs, with two buyer actions and three seller actions.}
\end{figure}

\begin{figure}
\centering
\begin{tikzpicture}[
x=1.65cm,
y=9cm,
scale=1.5,
font=\normalsize,
thick,
>=stealth',
dot/.style={
draw,
fill=white,
circle,
inner sep=0pt,
minimum size=4pt
}
]

\path[fill=gray!30] (0,0) -- (2.5,0.25) -- (3,0) -- cycle;

\path[fill=gray!70] (0,0) -- (1,{1/10}) -- (1,0) -- cycle;

\path[fill=blue,opacity=0.2]
(0,0) -- ({5/3},{1/6}) -- ({5/3},0) -- cycle;

\draw (0,0) -- (2.5,0.25) -- (3,0) -- cycle;
\draw[ultra thick] (2.5,0.25) -- (3,0);

\draw[dotted] (1,0) -- (1,{1/10});
\draw[dotted] ({5/3},0) -- ({5/3},{1/6});

\draw[->] (0,0) -- (0,0.34) node[left] {$V$};
\draw[->] (-0.25,0) -- (3.6,0) node[below] {$U$};
\node[below left] at (0,0) {$0$};
\draw (1,0) -- (1,-0.008);
\draw ({5/3},0) -- ({5/3},-0.008);
\node[below] at (1,-0.008) {$1$};
\node[below] at ({5/3},-0.008) {$\frac53$};

\node[dot] at (2.5,0.25) {};
\node at (2.5,0.285) {$(u_{2b},v_{2b})=(5/2,1/4)$};
\node at (3,-0.03) {$\overline U=3$};

\node at (2.6,0.035) {$\mathcal F$};
\node at (2.9,0.13) {$\mathcal L$};
\node at (0.65,0.21) {$\displaystyle\lim_{\gd\ra1}\mathcal E_\gd$};
\draw[thin,->] (0.7,0.18) -- (0.6,0.04);
\node[sgblue] at (3,0.21) {$\displaystyle\lim_{\gd\ra1}\mathcal P_\gd$};
\draw[sgblue,thin,->] (2.8,0.18) -- (1.4,0.075);
\end{tikzpicture}
\caption{Strict expansion and separation from the efficiency frontier, with $w_b=5$, $(p_{0b},p_{1b},p_{2b})=(1/4,1/2,3/4)$, $(c_{0b},c_{1b},c_{2b})=(0,1,5/2)$, and buyer payoffs $\overline v_b=1$ and $\underline v_b=-2$.}
\label{fig:ppe-pce-strict-separated}
\end{figure}

\section{Concluding remarks}

This paper has shown that mediation can eliminate the inefficiency that arises when buyers' myopic willingness to trade constrains the seller's dynamic incentives. I have studied a canonical moral-hazard setting in which gains from trade are constrained without mediation and shown that mediation can bring equilibrium payoffs arbitrarily close to every point on the efficiency frontier, even without type uncertainty. I have also shown that realizing these gains in full requires nonrecursive mediation, even as the seller becomes arbitrarily patient.

In my model, the mediator must overcome a filtering problem: each buyer uses the evolving public history and her private message to form beliefs about the seller's private history and hence his action. Tracking beliefs over private histories is a well-known obstacle to tractable analysis in repeated games. My ladder construction handles it by keeping buyers' beliefs over trade states constant conditional on a trade recommendation while giving the seller obedience incentives. The stage-game structure, especially binary outputs, helps make the construction tractable. Whether comparable gains are available in more general settings remains an important open question.

\pagebreak
\appendix

%
%

\begin{appendices}

\section{Preliminaries: effort incentives}
\label{sec:proofprelim}

For each trade $b>0$, let $d_{0b}:=0$ and define
\begin{align}\label{eq:cost-slopes}
d_{ab}:=\frac{c_{ab}-c_{(a-1)b}}{p_{ab}-p_{(a-1)b}},\qquad  a = 1,\dots,\overline a.
\end{align}
This is positive and nondecreasing in $a$ by convexity of $c_{ab}$ and concavity of $p_{ab}$ in $a$. 

In the no-mediation benchmark, in a typical trade $b>0$ on path in any PPE, let $U_q$ denote the seller's continuation payoff after output $q$ and define the scaled good-minus-bad continuation-payoff difference by
\begin{align*}
L := \frac{\gd(U_{\overline q}-U_{\underline q})}{1-\gd}.
\end{align*}
His payoff from effort $a$ in this trade is
\begin{align*}
&~ (1-\gd)(w_b-c_{ab})+\gd( p_{ab}U_{\overline q}+(1-p_{ab})U_{\underline q})\\
=&~ (1-\gd)w_b+\gd U_{\underline q}+(1-\gd)(p_{ab}L-c_{ab}).
\end{align*}
His optimal effort therefore maximizes $p_{ab}L-c_{ab}$.

\begin{lem*}\label{lem:optimala}
In this trade, effort $a$ is optimal---that is, the seller has no profitable deviation from $a$---if and only if
\begin{align}\label{eq:effort-incentives}
\begin{cases}
L\le d_{1b},    \quad &a=0,\\
d_{ab}\le L\le d_{(a+1)b},   \quad &0<a<\overline a,\\
L\ge d_{\overline a b},     \quad &a=\overline a.
\end{cases}
\end{align}
\end{lem*}

\begin{proof}[Proof of \cref{lem:optimala}]
Necessity of \eqref{eq:effort-incentives} follows by comparing $a$ with $a-1$ and $a+1$, whenever these adjacent efforts exist. Conversely, suppose \eqref{eq:effort-incentives} holds. For every $i<a$,
\begin{align*}
(p_{ab}L-c_{ab})-(p_{ib}L-c_{ib})
&=\sum_{j=i+1}^{a}(p_{jb}-p_{(j-1)b})(L-d_{jb})
\ge0,
\end{align*}
since $d_{jb}\le d_{ab}\le L$ for $j\le a$. For every $i>a$,
\begin{align*}
(p_{ab}L-c_{ab})-(p_{ib}L-c_{ib})
&=\sum_{j=a+1}^{i}(p_{jb}-p_{(j-1)b})(d_{jb}-L)
\ge0,
\end{align*}
since $d_{jb}\ge d_{a+1,b}\ge L$ for $j\ge a+1$. Thus $a$ is optimal. For every $a\ge1$, setting $L=d_{ab}$ satisfies \eqref{eq:effort-incentives} for effort $a$. Moreover, $(p_{ab}-p_{(a-1)b})L=c_{ab}-c_{(a-1)b}$, so $a-1$ yields the same payoff and is also optimal.
\end{proof}

\section{Proofs}
\label{sec:proofs}

\subsection{Proof of \cref{prop:Flbenchmark}}
\label{sec:proofbenchmark}

For each trade $b>0$, let $r_b:=\min\{a:v_{ab}\ge0\}$ be the lowest (pure) effort that makes the buyer willing to take trade $b$. For each effort $a>0$, define
\begin{align}\label{eq:kappa-ab}
\gk_{ab}:=(1-p_{ab})
\frac{c_{ab}-c_{(a-1)b}}
{p_{ab}-p_{(a-1)b}}>0,
\end{align}
and set $\gk_{0b}:=0$. Define also
\begin{align}\label{eq:gkppe}
\overline\gk_{\mathrm{PPE}}
:=\overline U-\max\left\{0,\max_{b \in B\setminus\{0\} }(u_{r_b b}-\gk_{r_b b})\right\}.
\end{align} 

I proceed in three steps.

\paragraph{Step 1.} I first establish the upper bound on the seller's PPE payoff. By definition,
\begin{align*}
\overline U-\overline\gk_{\mathrm{PPE}}
=\max\left\{0,\max_{b \in B \setminus \{0\}}(u_{r_b b}-\gk_{r_b b})\right\}.
\end{align*}
Fix $\gd$ and write $\overline U_{\mathrm{PPE}}:=\sup_{(U,V)\in\mathcal E_\gd}U$. Every continuation is a PPE, so its seller payoff is at most $\overline U_{\mathrm{PPE}}$. If the buyer chooses trade $b$ with positive probability, some effort $a\ge r_b$ must occur with positive probability; otherwise she would prefer rejection. The seller is indifferent between all effort levels to which he assigns positive probability. Comparing effort $a$ with $a-1$ gives
\begin{align*}
U_{\overline q}-U_{\underline q}\ge\frac{1-\gd}{\gd}d_{ab}.
\end{align*}
The seller's continuation payoff conditional on $b$ therefore equals his continuation payoff from effort $a$ and satisfies
\begin{align*}
(1-\gd)u_{ab}+\gd U_{\overline q}
-\gd(1-p_{ab})(U_{\overline q}-U_{\underline q})
&\le(1-\gd)(u_{ab}-\gk_{ab})+\gd\overline U_{\mathrm{PPE}}\\
&\le(1-\gd)(u_{r_b b}-\gk_{r_b b})+\gd\overline U_{\mathrm{PPE}}\\
&\le(1-\gd)(\overline U-\overline\gk_{\mathrm{PPE}})
+\gd\overline U_{\mathrm{PPE}}.
\end{align*}
The second inequality holds because $u_{ab}-\gk_{ab}$ is nonincreasing in $a \ge r_b$:
\begin{align*}
(c_{a+1,b}+\gk_{a+1,b})-(c_{ab}+\gk_{ab})
=(1-p_{ab})(d_{a+1,b}-d_{ab})\ge0,
\qquad a = 1, \dots, \overline a-1.
\end{align*}
Rejection gives the seller at most $\gd\overline U_{\mathrm{PPE}}$. Therefore, averaging over buyer actions (and public randomization outcomes), then taking the supremum, gives $\overline U_{\mathrm{PPE}}
\le(1-\gd)(\overline U-\overline\gk_{\mathrm{PPE}})
+\gd\overline U_{\mathrm{PPE}}$, and hence $\overline U_{\mathrm{PPE}}\le\overline U-\overline\gk_{\mathrm{PPE}}$. Moreover, $u_{r_b b}-\gk_{r_b b}<u_{r_b b}\le\overline U$. Since there are finitely many trade actions and $\overline U>0$, $\overline\gk_{\mathrm{PPE}}>0$.

If $\overline U-\overline\gk_{\mathrm{PPE}}=0$, the only feasible payoff satisfying the bound is $(0,0)$, because every trade gives the seller a positive payoff. Permanent rejection attains this vector for every $\gd$. Henceforth suppose $\overline U-\overline\gk_{\mathrm{PPE}}>0$.

\paragraph{Step 2.} I show that every $(U,0)$ with $U\in[0,\overline U-\overline\gk_{\mathrm{PPE}}]$ is a PPE payoff for sufficiently high $\gd$. Since permanent rejection (with shirking in unexpected trades) attains $(0,0)$ and public randomization is available, it suffices to construct a PPE attaining $(\overline U-\overline\gk_{\mathrm{PPE}},0)$ for sufficiently high $\gd$. A two-state PPE attains this latter payoff vector for sufficiently high $\gd$; call one state the trade state and call the other state the rejection state. Choose $b$ with $u_{r_b b}-\gk_{r_b b}=\overline U-\overline\gk_{\mathrm{PPE}}$; such $b$ exists because $\overline U-\overline\gk_{\mathrm{PPE}}>0$. In the trade state, the buyer chooses $b$, and the seller chooses effort $r_b$ with probability $-v_{(r_b-1)b}/(v_{r_b b}-v_{(r_b-1)b})$ and effort $r_b-1$ otherwise. A good output leaves the state unchanged. After a bad output, public randomization at the beginning of the next period selects the rejection state with probability
\begin{align*}
\frac{(1-\gd)d_{r_b b}}{\gd(\overline U-\overline\gk_{\mathrm{PPE}})},
\end{align*}
and otherwise leaves the state unchanged. This probability lies in $(0,1]$ for sufficiently high $\gd$. Any buyer deviation leads to rejection from the next period, with shirking in an unexpected trade. The rejection state is absorbing and prescribes rejection and shirking in every unexpected trade.

The seller's state payoffs are $\overline U-\overline\gk_{\mathrm{PPE}}$ and zero by direct calculation. In the trade state, the good-minus-bad continuation-payoff difference, $U_{\overline q} - U_{\underline q}$, equals $(1-\gd)d_{r_b b}/\gd$, so both prescribed efforts are optimal by \cref{lem:optimala} (in \cref{sec:proofprelim}). The buyer is indifferent between trade $b$ and rejection, since
\begin{align*}
\frac{-v_{(r_b-1)b}}{v_{r_b b}-v_{(r_b-1)b}}v_{r_b b}
+\frac{v_{r_b b}}{v_{r_b b}-v_{(r_b-1)b}}v_{(r_b-1)b}
=0.
\end{align*}
Every other trade induces shirking and gives a negative payoff. After any buyer deviation and in the rejection state, shirking is optimal because future play is permanent rejection. Consequently, the strategies form a PPE.

\paragraph{Step 3.} Fix $(U,V)\in\operatorname{int}\mathcal F$ with $0<U<\overline U-\overline\gk_{\mathrm{PPE}}$. Choose $\widetilde V>V$ with $(U,\widetilde V)\in\operatorname{int}\mathcal F$. I construct PPE payoffs in which the seller's payoff is $U$ for all sufficiently high $\gd$ and the average buyer's payoff converges to $\widetilde V$ as $\gd \ra 1$. I then use public randomization between this equilibrium and the PPE attaining $(U,0)$ to achieve a PPE attaining $(U,V)$ for all sufficiently high $\gd$.

The PPE has a main phase and a residual phase. Let $U'_t$ be the seller's continuation payoff at a typical public history $h_t$ in this PPE. Set $U'_0=U$. In each main-phase period, public randomization selects among rejection, trade $b$ with pure effort $a$ satisfying $v_{ab}\ge0$, or trade $b$ with the mixture of $r_b-1$ and $r_b$ that gives the buyer payoff zero, so that the expected current payoffs are $(U,\widetilde V)$.

The next-period continuation payoff after each output is specified as follows. If public randomization prescribed trade $b$ with pure effort $a$, set
\begin{align}\label{eq:y}
U'_{t+1}=
\begin{cases}
\dfrac{U'_t-(1-\gd)(u_{ab}-\gk_{ab})}{\gd},
&\text{after }\overline q,\\[0.8em]
\dfrac{U'_t-(1-\gd)(u_{ab}-\gk_{ab})}{\gd}
-\dfrac{1-\gd}{\gd}d_{ab},
&\text{after }\underline q.
\end{cases}
\end{align}
If public randomization prescribed the mixed-action prescription, use the same formulas with $a=r_b$. If public randomization prescribed rejection, set $U'_{t+1}=U'_t/\gd$. A buyer deviation induces shirking if trade occurs and permanent rejection from the next period. By construction, every effort prescribed with positive probability satisfies \eqref{eq:effort-incentives}, and the seller's promise-keeping condition holds: $U'_t = (1-\gd) \bE[u_t|h_t] + \gd \bE[U'_{t+1} | h_t]$.

The main phase continues while the seller's continuation payoff remains in a fixed interval around $U$. Specifically, choose $\eta>0$ with $2\eta<\min\{U,\overline U-\overline\gk_{\mathrm{PPE}}-U\}$, and the main phase continues while
\begin{align}\label{eq:interval}
U'_t\in[U-\eta,U+\eta].
\end{align}
Let $R:=\max_{b>0}w_b+\max_{b>0}d_{\overline a b}$. Every continuation-payoff update from this interval satisfies $|U'_{t+1}-U'_t|\le(1-\gd)R/\gd<\eta$ for sufficiently high $\gd$. Thus every next-period continuation payoff lies in $[U-2\eta,U+2\eta]\subset(0,\overline U-\overline\gk_{\mathrm{PPE}})$. If $U'_{t+1}$ leaves \eqref{eq:interval}, the residual phase begins in period $t+1$ with the PPE attaining $(U'_{t+1},0)$ from Step 2.

The residual phase has vanishing expected discounted weight. Let $\tau$ be its first period, with $\gd^\infty:=0$. I show that
\begin{align}\label{eq:exit}
\bE[\gd^\tau]\le\frac{R}{\eta}\sqrt{1-\gd}\ra 0.
\end{align}
For $t < \tau$, $U'_t=(1-\gd)U+\gd\bE[U'_{t+1}\mid h_t]$. Hence
\begin{align}\label{eq:identity}
M_t:=\gd^{t\wedge\tau}(U-U'_{t\wedge\tau})
\end{align}
is a bounded martingale with $M_0=0$. Its limit is $M_\infty=\gd^\tau(U-U'_\tau)$ if $\tau<\infty$ and zero otherwise. Moreover, by construction, $|M_{t+1}-M_t|\le(1-\gd)\gd^tR$. Also, for $s<t$, the martingale property implies $\bE[(M_{s+1}-M_s)(M_{t+1}-M_t)]=0$. Therefore
\begin{align}\label{eq:variance}
\bE[M_\infty^2] 
=\sum_{t=0}^\infty \bE[(M_{t+1}-M_t)^2]
\le(1-\gd)^2R^2\sum_{t=0}^\infty\gd^{2t}
\le(1-\gd)R^2.
\end{align}
On $\{\tau<\infty\}$, $|U-U'_\tau|>\eta$, so $\gd^\tau\le|M_\infty|/\eta$ on every path. By Cauchy--Schwarz,
\begin{align}\label{eq:maximal}
\bE[\gd^\tau]
\le\frac{\bE[|M_\infty|]}{\eta}
\le\frac{\sqrt{\bE[M_\infty^2]}}{\eta}
\le\frac{R}{\eta}\sqrt{1-\gd},
\end{align}
which proves \eqref{eq:exit}.

The constructed PPE has seller payoff $U$ and average-buyer payoff converging to $\widetilde V$ as $\gd \ra 1$. Each main-phase period gives the buyer expected payoff $\widetilde V$ conditional on $h_t$, and the residual phase gives zero. Thus its payoff vector is $(U,V_\gd)$, where
\begin{align*}
V_\gd:=\widetilde V(1-\bE[\gd^\tau]) \ra \widetilde V, \quad \text{ as } \gd \ra 1.
\end{align*}
For sufficiently high $\gd$, $V_\gd>V$. Public randomization between this PPE and the PPE attaining $(U,0)$, with probability $V/V_\gd$ on the former, attains $(U,V)$. The desired Hausdorff convergence follows: every PPE payoff lies in the set on the right of \eqref{eq:limitPPE} by Step 1, and since the interior targets of Step 3 are dense in that set and the set is compact, finitely many of them, each attained for sufficiently high $\gd$, lie within any $\ve>0$ of every point of the set.


\subsection{Proof of \cref{thm:payoffset}}

\cref{lem:construction} below, which I prove momentarily, is essential:

\begin{lem*}\label{lem:construction}
For every trade $b>0$, every $(U,V)\in\mathcal L_b$, and every neighborhood of $(U,V)$, there exists $\gd_0<1$ such that the neighborhood contains an SSCE payoff vector at every $\gd\in(\gd_0,1)$.
\end{lem*}

Since $\mathcal F=\operatorname{co}(\{(0,0)\}\cup\bigcup_{b>0}\mathcal L_b)$, \cref{lem:construction} implies that each nonzero vertex of $\mathcal F$ lies on some $\mathcal L_b$ and can be approximated by SSCE payoff vectors for all sufficiently high $\gd$. Permanent rejection, with shirking in unexpected trades, attains an SSCE with payoff $(0,0)$. Fix any target payoff vector in the interior of $\mathcal F$. Because it lies in the interior, it remains in the convex hull of the vertices of $\mathcal F$ when each vertex is replaced by a sufficiently close SSCE payoff vector, which exists for every vertex by \cref{lem:construction} once the seller is sufficiently patient. The mediator therefore attains the target by randomizing at the outset over these SSCEs with the corresponding weights and revealing the outcome to every player. Hausdorff convergence then follows from the compactness of $\mathcal F$ as in the proof of \cref{prop:Flbenchmark}. This proves \cref{thm:payoffset}. 

\begin{proof}[Proof of \cref{lem:construction}]
Fix $b>0$. Choose $a^*$ and $\gl\in(0,1)$ with $\pi=(1-\gl)p_{(a^*-1)b}+\gl p_{a^*b}$ and $V^*_{b,\pi}>0$. Recall that $c_b^*:=(1-\gl)c_{(a^*-1)b}+\gl c_{a^*b}=w_b-U^*_{b,\pi}$. These targets are dense in $\mathcal L_b$, as $(a^*, \gl)$ vary. It therefore suffices to construct SSCE payoffs approaching $(U^*_{b,\pi},V^*_{b,\pi})$ as $\ve\ra0$, with the seller sufficiently patient for each $\ve$. Let $\ve\in(0,U^*_{b,\pi})$ be sufficiently small and generic, chosen below. The device sends action recommendations, with the seller's recommendation contingent on the buyer's action. A strategy profile is obedient if each player follows these recommendations at every private history.

\paragraph{Hidden states.}
The device uses trade states $1,\ldots,N$ and an absorbing no-trade state $0$. I choose $N\ge3$ and efforts $a_1=\overline a\ge\cdots\ge a_N$ with $p_{a_Nb}<\pi<p_{\overline a b}$ below. Initially, the state is drawn from a distribution $\rho$ over the trade states. At trade state $n$, the device tells the seller, but not the buyer, the state, recommends $b$ to the buyer, and prescribes effort $a_n$ after $b$ and shirking after every other trade. At state $0$, it recommends rejection and shirking in every unexpected trade. Every buyer deviation sends the next state to $0$.

Under the local rule, an interior state moves up with probability $1-\pi$ after a good output and down with probability $\pi$ after a bad output. At the top state, a bad output moves the state down with probability $(\pi-p_{a_Nb})/(1-p_{a_Nb})$. At the bottom state, a good output moves the state up with probability $(p_{\overline a b}-\pi)/p_{\overline a b}$. In all other cases, the state stays unchanged. After each prescribed trade, the device follows this local rule with probability $1-\xi$ and redraws from $\rho$ with probability $\xi$. After a bad output at state $1$, probability $\eta$ of a redraw is instead assigned to rejection. I choose $0<\eta<\xi<1$ below.

Given the effort profile, choose $\rho_n>0$ with $\sum_n\rho_n=1$ satisfying
\begin{align}\label{eq:SSCEadjacentstates}
\rho_1(p_{\overline a b}-\pi)
&=\rho_2\pi(1-p_{a_2b}),\nonumber\\
\rho_np_{a_nb}(1-\pi)
&=\rho_{n+1}\pi(1-p_{a_{n+1}b}),
\qquad n=2,\dots,N-2,\\
\rho_{N-1}p_{a_{N-1}b}(1-\pi)
&=\rho_N(\pi-p_{a_Nb}).\nonumber
\end{align}
These equations uniquely determine $\rho$, and $\rho$ is stationary under the local rule. The expected change in the state index at state $n$ is $p_{a_nb}-\pi$. Stationarity therefore gives $\sum_n\rho_np_{a_nb}=\pi$.

\paragraph{Buyer incentives.}
Suppose the current trade state has distribution $\rho$. Equations \eqref{eq:SSCEadjacentstates} imply that, under the local rule, the joint probabilities of a good or bad output and next state $i$ are $\pi\rho_i$ and $(1-\pi)\rho_i$, respectively. Redrawing from $\rho$ preserves these probabilities. Rejection removes probability $\rho_1(1-p_{\overline a b})\eta\rho_i$ from each next trade state $i$ after a bad output. Thus,
\begin{align*}
\Pr(\overline q,\text{ next state }i)&=\pi\rho_i,\nonumber\\
\Pr(\underline q,\text{ next state }i)
&=\bigl[1-\pi-\rho_1(1-p_{\overline a b})\eta\bigr]\rho_i.
\end{align*}
Both expressions are proportional to $\rho_i$. Since the initial state has distribution $\rho$, induction shows that every buyer recommended trade $b$ has posterior $\rho$. Her expected payoff is therefore $V^*_{b,\pi}>0$, compared with zero from rejection and $v_{0b'}<0$ from any other trade $b'$. A rejection recommendation reveals state $0$, where every trade induces shirking. Thus every buyer strictly prefers to obey.

\paragraph{Prescribed efforts.}
I choose the efforts and gaps from the top down. As in the sketch, $g_n$ will be the limit of $\gd(U_{n+1}-U_n)/(1-\gd)$ as $(\xi,\gd)\ra(0,1)$, where $U_n$ is the seller's continuation payoff in trade state $n$. I construct the gaps first and verify their payoff interpretation below. Set $g_N:=0$. Recall
\begin{align*}
J_{ab}:=c_{ab}+(1-p_{ab})
\frac{c_{ab}-c_{(a-1)b}}{p_{ab}-p_{(a-1)b}},
\qquad a=1,\dots,\overline a,
\end{align*}
with $J_{0b}=0$, and set $J_{(\overline a+1)b}:=\infty$. These values are nondecreasing in $a$. Restrict $\ve>0$ so that $c_b^*+\ve<\min\{c_{\overline a b},J_{a^*b}\}$. This is possible because $c_b^*<c_{a^*b}\le\min\{c_{\overline a b},J_{a^*b}\}$.

At the top state, prescribe effort $a_N:=\max\{a\in A:J_{ab}<c_b^*+\ve\}$, where $N$ will be determined below. For all sufficiently small $\ve>0$, this effort is constant and $J_{a_Nb}<c_b^*+\ve<J_{(a_N+1)b}$. Since $c_b^*+\ve<J_{a^*b}$, $a_N<a^*$ and hence $p_{a_Nb}<\pi$. Set
\begin{align}\label{eq:SSCEtopgap}
g_{N-1}:=\frac{c_b^*+\ve-c_{a_Nb}}{\pi-p_{a_Nb}}>0.
\end{align}
Then $w_b-c_{a_Nb}-(\pi-p_{a_Nb})g_{N-1}=U^*_{b,\pi}-\ve$, as required at the top state in the joint limit. By \cref{lem:optimala}, the strict inequalities involving $J_{a_Nb}$ and $J_{(a_N+1)b}$ make $a_N$ the unique maximizer of $p_{ab}(\pi-p_{a_Nb})g_{N-1}/(1-p_{a_Nb})-c_{ab}$ over $a\in A$.

At each interior state $n$, given $g_n$, prescribe effort $a_n:=\max\{a\in A:J_{ab}<c_b^*+\ve+(1-\pi)g_n\}$, as in \eqref{eq:selection}. Set
\begin{align}\label{eq:SSCEgapmap}
g_{n-1}:=\frac{p_{a_nb}(1-\pi)g_n+c_b^*+\ve-c_{a_nb}}{\pi(1-p_{a_nb})}.
\end{align}
This gives $w_b-c_{a_nb}+p_{a_nb}(1-\pi)g_n-\pi(1-p_{a_nb})g_{n-1}=U^*_{b,\pi}-\ve$. Moreover,
\begin{align*}
(1-\pi)g_n+\pi g_{n-1}
=\frac{c_b^*+\ve+(1-\pi)g_n-c_{a_nb}}{1-p_{a_nb}}.
\end{align*}
Since $J_{a_nb}<c_b^*+\ve+(1-\pi)g_n\le J_{(a_n+1)b}$, \cref{lem:optimala} implies that $a_n$ maximizes $p_{ab}[(1-\pi)g_n+\pi g_{n-1}]-c_{ab}$ over $a\in A$. The maximizer is unique whenever the upper inequality is strict.

Comparing the prescribed effort with the target mixture gives
\begin{align*}
\pi(1-\pi)(g_{n-1}-g_n)
&=c_b^*+\ve-c_{a_nb}
+(p_{a_nb}-\pi)[(1-\pi)g_n+\pi g_{n-1}]\nonumber\\
&\ge\ve.
\end{align*}
Thus $g_{n-1}\ge g_n+\ve/[\pi(1-\pi)]>g_n$. Every effort satisfying the selection inequality at state $n$ also satisfies it at the next interior state down, so $a_{n-1}\ge a_n$. Also, $a_{N-1}\ge a_N$ because $g_{N-1}>0$.

After at least one interior step, stop when the newly computed gap satisfies
\begin{align*}
w_b+(p_{\overline a b}-\pi)g_{n-1}-J_{\overline a b}
>U^*_{b,\pi}-\ve.
\end{align*}
The next lower state becomes the bottom state, where I prescribe $\overline a$. This occurs after finitely many steps because the gaps increase by at least $\ve/[\pi(1-\pi)]$ at each interior step and $p_{\overline a b}\ge p_{a^*b}>\pi$. Number the completed ladder $1,\dots,N$ from bottom to top. Then $N\ge3$, $a_1=\overline a\ge\cdots\ge a_N$, and the bottom gap is $g_1$. With this numbering, the stopping condition is \eqref{eq:stoppingc}.

I show that for generic $\ve$, the seller's obedience incentives are strict at each trade state. For each effort $a\in A$, the ratio
\begin{align*}
\frac{p_{ab}(1-\pi)g_n+c_b^*+\ve-c_{ab}}{\pi(1-p_{ab})}
\end{align*}
is the gap at state $n-1$ that promise keeping at state $n$ would require if effort $a$ were prescribed there, given $g_n$; by \eqref{eq:SSCEgapmap}, $g_{n-1}$ is this ratio at $a=a_n$. Rearranging \eqref{eq:SSCEgapmap}, the ratio at $a$ is at least $g_{n-1}$ if and only if
\begin{align*}
p_{ab}[(1-\pi)g_n+\pi g_{n-1}]-c_{ab}
\ge p_{a_nb}[(1-\pi)g_n+\pi g_{n-1}]-c_{a_nb}.
\end{align*}
Since $a_n$ is a maximizer, the ratio at every $a\in A$ is at most $g_{n-1}$, with equality at $a=a_n$. Therefore
\begin{align}\label{eq:SSCEenvelope}
g_{n-1}
=\max_{a\in A}
\frac{p_{ab}(1-\pi)g_n+c_b^*+\ve-c_{ab}}{\pi(1-p_{ab})}.
\end{align}
The right side is continuous and strictly increasing in $\ve$ and $g_n$. For all sufficiently small $\ve>0$, $a_N$ is constant, so the initial gap in \eqref{eq:SSCEtopgap} is strictly increasing in $\ve$. Repeatedly applying \eqref{eq:SSCEenvelope} gives the same conclusion after any fixed number of steps from the top. At each such step, each equality $c_b^*+\ve+(1-\pi)g_n=J_{ab}$ therefore excludes at most one value of $\ve$. There are finitely many efforts and countably many possible steps, so at most countably many values of $\ve$ are excluded. All the preceding effort maximizers are therefore unique, given the generic $\ve$.

\paragraph{Seller incentives and the redraw probability.}
I choose the redraw probability $\xi$ so that a sufficiently patient seller has strict obedience incentives at every trade state. Since the initial state is drawn from $\rho$, attaining initial seller payoff $U^*_{b,\pi}-\ve$ requires
\begin{align}\label{eq:prelim}
\sum_n\rho_nU_n=U^*_{b,\pi}-\ve.
\end{align}
The seller's continuation payoff at each trade state except the bottom satisfies:
\begin{align}\label{eq:SSCEphi}
U_n=(1-\gd)u_{a_nb}
+\gd[p_{a_nb}U_{n,\overline q}+(1-p_{a_nb})U_{n,\underline q}],
\qquad n=2,\dots,N.
\end{align}
Here, for $n=2,\dots,N$, $U_{n,q}$ is the expected next-period payoff after output $q$, using the local transitions with probability $1-\xi$ and a redraw from $\rho$ with probability $\xi$. Using \eqref{eq:prelim}, rewrite \eqref{eq:SSCEphi} as a linear system in the scaled adjacent differences $\gd(U_n-U_{n-1})/(1-\gd)$, with coefficients continuous near $(\xi,\gd)=(0,1)$. In the limit as $(\xi, \gd) \ra (0,1)$, this system reduces to \eqref{eq:SSCEtopgap} and \eqref{eq:SSCEgapmap}, which uniquely determine the gaps from the top down. Its solution therefore remains unique and continuous nearby. Consequently,
\begin{align}\label{eq:SSCEactualIC}
\lim_{(\xi,\gd)\ra(0,1)}
\frac{\gd(U_n-U_{n-1})}{1-\gd}=g_{n-1},
\qquad n=2,\dots,N.
\end{align}
At the top state, the scaled good-minus-bad continuation payoff difference is
\begin{align*}
\lim_{(\xi,\gd)\ra(0,1)} \frac{\gd}{1-\gd}(U_{N,\overline q}-U_{N,\underline q})
&=\lim_{(\xi,\gd)\ra(0,1)} (1-\xi)\frac{\pi-p_{a_Nb}}{1-p_{a_Nb}}
\frac{\gd(U_N-U_{N-1})}{1-\gd}\\
&=
\frac{\pi-p_{a_Nb}}{1-p_{a_Nb}}g_{N-1}.
\end{align*}
Similarly, at each interior state $n=2,\dots,N-1$,
\begin{align*}
\lim_{(\xi,\gd)\ra(0,1)} \frac{\gd}{1-\gd}(U_{n,\overline q}-U_{n,\underline q})
&= \lim_{(\xi,\gd)\ra(0,1)} (1-\xi)\left[
(1-\pi)\frac{\gd(U_{n+1}-U_n)}{1-\gd}
+\pi\frac{\gd(U_n-U_{n-1})}{1-\gd}
\right]\\
&= (1-\pi)g_n+\pi g_{n-1}.
\end{align*}
These are the same scaled good-minus-bad continuation-payoff differences for which the prescribed efforts were shown above to be strictly optimal in the limit as $(\xi, \gd) \ra (0,1)$. Finally, using the choice of $\eta$ in \eqref{eq:SSCErejection} and the identity \eqref{eq:SSCEadjacentpayoffs}, both established below, at the bottom trade state,
\begin{align*}
\lim_{(\xi,\gd)\ra(0,1)}
\frac{\gd(U_{1,\overline q}-U_{1,\underline q})}{1-\gd}
&=\frac{p_{\overline a b}-\pi}{p_{\overline a b}}g_1
+\frac{c_b^*+\ve-c_{\overline a b}+(p_{\overline a b}-\pi)g_1}{1-p_{\overline a b}}\nonumber\\
&>\frac{c_{\overline a b}-c_{(\overline a-1)b}}{p_{\overline a b}-p_{(\overline a-1)b}},
\end{align*}
where the inequality follows from the stopping condition. Therefore I can fix $\xi>0$ sufficiently small so that every prescribed effort remains strictly optimal for all sufficiently high $\gd<1$, once the rejection probability meets the stated requirement.

\paragraph{Rejection probability and seller payoffs.}
I first show that the seller's expected current trade payoff under $\rho$ exceeds $U^*_{b,\pi}-\ve$. Multiplying \eqref{eq:SSCEtopgap} by $\pi-p_{a_Nb}$ and \eqref{eq:SSCEgapmap} by $\pi(1-p_{a_nb})$, then rearranging, weighting by $\rho_n$, and summing over $n=2,\dots,N$, gives, using \eqref{eq:SSCEadjacentstates}, $\sum_{n=2}^N\rho_n(c_b^*+\ve-c_{a_nb}) =\rho_1(p_{\overline a b}-\pi)g_1$. Consequently,
\begin{align}\nonumber
\sum_n\rho_nu_{a_nb}-(U^*_{b,\pi}-\ve)
&=\rho_1[c_b^*+\ve-c_{\overline a b}+(p_{\overline a b}-\pi)g_1] \\
&>0, \label{eq:SSCEadjacentpayoffs}
\end{align}
where positivity follows from the stopping condition and $J_{\overline a b}>c_{\overline a b}$. Averaging the payoff equations over all trade states under $\rho$, subject to \eqref{eq:prelim}, requires
\begin{align*}
U^*_{b,\pi}-\ve
&=(1-\gd)\sum_n\rho_nu_{a_nb}
+\gd [1-\rho_1(1-p_{\overline a b})\eta ](U^*_{b,\pi}-\ve).
\end{align*}
This gives the required condition on the rejection probability $\eta$. Therefore, set
\begin{align}\label{eq:SSCErejection}
\eta
=\frac{(1-\gd)[\sum_n\rho_nu_{a_nb}-(U^*_{b,\pi}-\ve)]}
{\gd(U^*_{b,\pi}-\ve)\rho_1(1-p_{\overline a b})}.
\end{align}
This is positive by \eqref{eq:SSCEadjacentpayoffs} and tends to zero as $\gd\ra1$. Choose $\gd_0<1$ so that $\eta<\xi$ and all the preceding effort comparisons are strict whenever $\gd\in(\gd_0,1)$.

The seller's continuation payoff in each trade state $n$ under obedience satisfies
\begin{align*}
U_n=(1-\gd)u_{a_nb}
+\gd[p_{a_nb}U_{n,\overline q}+(1-p_{a_nb})U_{n,\underline q}],
\qquad n=1,\dots,N,
\end{align*}
with payoff zero in the no-trade state. Averaging under $\rho$ gives 
\begin{align*}
\sum_n\rho_nU_n=(1-\gd)\sum_n\rho_nu_{a_nb}+\gd[1-\rho_1(1-p_{\overline a b})\eta]\sum_n\rho_nU_n.    
\end{align*}
By \eqref{eq:SSCErejection}, this implies \eqref{eq:prelim}. The seller's continuation payoffs under the completed device therefore satisfy \eqref{eq:prelim} and \eqref{eq:SSCEphi} and, by uniqueness, equal the $U_n$ used above. This verifies the seller's strict effort incentives and the interpretation of the gaps in \eqref{eq:SSCEactualIC}, with the joint limit taken through parameter values satisfying $0<\eta<\xi$. For the fixed $\xi$ chosen above, the scaled adjacent differences $\gd(U_n-U_{n-1})/(1-\gd)$ remain bounded as $\gd\ra1$, so the differences between the seller's continuation payoffs vanish. Since their weighted average payoff under $\rho$ is $U^*_{b,\pi}-\ve$, every $U_n$ converges to $U^*_{b,\pi}-\ve$ as $\gd \ra 1$.

\paragraph{Sequential rationality and consistency.}
The seller knows the state, so the strict effort comparisons apply after every private history with a prescribed trade. In state $0$ and after a buyer deviation, shirking uniquely maximizes his current payoff and cannot affect future play. Since payoffs are bounded and discounted, the one-shot deviation principle rules out every seller deviation. Together with the buyer incentives shown above, this establishes sequential rationality.

To verify consistency, perturb the prescribed strategies by assigning equal, vanishing probabilities to all actions other than the recommended action. These completely mixed strategies converge to obedience, and their Bayesian beliefs converge at every finite feasible history. The limiting beliefs agree with the buyer posteriors used above: every feasible buyer history with a trade recommendation has positive probability under obedience, while a rejection recommendation reveals state $0$. Thus the construction is an SSCE.

\paragraph{Payoff convergence.}
The seller's initial payoff is $U=U^*_{b,\pi}-\ve$. To bound the average buyer's payoff, I first show that the seller's expected current payoff conditional on trade is at most $U^*_{b,\pi}$. The nondecreasing ratios in \eqref{eq:cost-slopes} imply 
\begin{align*}
c_{ab}\ge c_{(a^*-1)b} + \dfrac{c_{a^*b}-c_{(a^*-1)b}}{p_{a^*b}-p_{(a^*-1)b}} (p_{ab}-p_{(a^*-1)b})
\end{align*}
for every $a$. Averaging and using $\sum_n\rho_np_{a_nb}=\pi$ gives $\sum_n\rho_nc_{a_nb}\ge c_b^*$. Consequently, $U^*_{b,\pi}-\ve<\sum_n\rho_nu_{a_nb}\le U^*_{b,\pi}$.

Conditional on a trade recommendation, the expected current payoffs are $(\sum_n\rho_nu_{a_nb},V^*_{b,\pi})$. Rejection gives both players zero. Hence $\bE[v_t]=[V^*_{b,\pi}/\sum_n\rho_nu_{a_nb}]\bE[u_t]$ in every period. Multiplying by $(1-\gd)\gd^t$ and summing gives $V=[(U^*_{b,\pi}-\ve)/\sum_n\rho_nu_{a_nb}]V^*_{b,\pi}$, so $(1-\ve/U^*_{b,\pi})V^*_{b,\pi}\le V\le V^*_{b,\pi}$. For each $\ve$ chosen above, these payoffs are attained for every sufficiently high $\gd$. Letting $\ve\ra0$ through such values approaches the target.
\end{proof}

\subsection{Proof of \cref{prop:uniformseparation}}
\label{sec:proofuniformseparation}

The proof follows a similar structure to that of \cref{prop:Flbenchmark}. For any distribution $\zeta \in \Delta(A)$, let $\zeta_a$ denote the probability assigned to $a \in A$. Define
\begin{align*}
\overline\gk_{\mathrm{PCE}}
:=\overline U-\max\left\{0,\,
\max_{b\in B\setminus\{0\}}
\max_{\substack{\zeta\in\Delta(A)\\
\sum_{a=0}^{\overline a}\zeta_a v_{ab}\ge0}}
\sum_{a=0}^{\overline a}\zeta_a(u_{ab}-\gk_{ab})
\right\}.
\end{align*}

By definition of \eqref{eq:gkppe}, $\overline\gk_{\mathrm{PCE}}\le\overline\gk_{\mathrm{PPE}}$. Recall the definition of $\gk_{ab}$ from \eqref{eq:kappa-ab}. I proceed in three steps.

\paragraph{Step 1.} I first establish the upper bound on the seller's PCE payoff. By definition,
\begin{align}\label{eq:many-PCEcap}
\overline U-\overline\gk_{\mathrm{PCE}}
=\max\left\{0,\max_{b>0}
\max_{\substack{\zeta\in\Delta(A)\\ \sum_a\zeta_av_{ab}\ge0}}
\sum_a\zeta_a(u_{ab}-\gk_{ab})\right\}.
\end{align}
Fix $\gd$ and write $\overline U_{\mathrm{PCE}}:=\sup_{(U,V)\in\mathcal P_\gd}U$. At any mediator history before the current recommendations, let $\mu_{ab}$ be the probability of recommending trade $b$ and effort $a$ in that trade, and let $\mu_0$ be the probability of recommending rejection. Let $U_{ab,q}$ be the average seller continuation payoff conditional on this pair and output $q$. Every continuation is a PCE, so $U_{ab,q}\le\overline U_{\mathrm{PCE}}$. The obedience constraints for any such recommendation profile are
\begin{align}
\mu_{ab}>0 \quad \implies \quad U_{ab,\overline q}-U_{ab,\underline q}
&\ge\frac{1-\gd}{\gd}d_{ab},
&&a = 1,\dots,\overline a,\label{eq:recursiveIC1}\\
\sum_a\mu_{ab}v_{ab}&\ge0, 
&&b=1,\dots, \overline b.\label{eq:recursiveIC2}
\end{align}
The seller's payoff conditional on recommendation of trade $b$ and effort $a$ in trade $b$ (and shirking otherwise) is therefore
\begin{align*}
(1-\gd)u_{ab}
+\gd ( p_{ab}U_{ab,\overline q}+(1-p_{ab})U_{ab,\underline q} ) \le (1-\gd)(u_{ab}-\gk_{ab})+\gd\overline U_{\mathrm{PCE}}.
\end{align*}
This upper bound also holds for $a=0$, since $\gk_{0b}=0$. Rejection gives at most $\gd\overline U_{\mathrm{PCE}}$. Therefore the above upper bound and \eqref{eq:many-PCEcap} together imply $\overline U_{\mathrm{PCE}}
\le(1-\gd)(\overline U-\overline\gk_{\mathrm{PCE}})
+\gd\overline U_{\mathrm{PCE}}$, and hence $\overline U_{\mathrm{PCE}}\le\overline U-\overline\gk_{\mathrm{PCE}}$.

The loss $\overline\gk_{\mathrm{PCE}}$ is strictly positive. For each $b$, the maximum over $\zeta$ in \eqref{eq:many-PCEcap} exists because the constraint set is nonempty and compact. Every mixture satisfying buyer participation assigns positive probability to positive effort, since $v_{0b}<0$. Thus, for every $\zeta\in\Delta(A)$ satisfying $\sum_a\zeta_av_{ab}\ge0$, $\sum_a\zeta_a(u_{ab}-\gk_{ab})
<\sum_a\zeta_au_{ab}\le\overline U$. Since each maximum is attained, there are finitely many trades, and $\overline U>0$, \eqref{eq:many-PCEcap} implies $\overline\gk_{\mathrm{PCE}}>0$.

If $\overline U-\overline\gk_{\mathrm{PCE}}=0$, the only feasible payoff satisfying the bound is $(0,0)$, attained by permanent rejection for every $\gd$. Henceforth suppose $\overline U-\overline\gk_{\mathrm{PCE}}>0$.

\paragraph{Step 2.} I show that every $(U,0)$ with $U\in[0,\overline U-\overline\gk_{\mathrm{PCE}}]$ is a PCE payoff for sufficiently high $\gd$. Since $(0,0)$ is attained by a PCE with permanent rejection and shirking in any unexpected trade and the mediator's public randomization over PCEs at the outset attains  a PCE, it suffices to construct a PCE attaining $(\overline U-\overline\gk_{\mathrm{PCE}},0)$. Choose $b$ and $\zeta$ attaining \eqref{eq:many-PCEcap}. This mixture gives the buyer zero: otherwise, mixing it with effort $0$ until her payoff becomes zero would strictly increase the objective, since $u_{0b}-\gk_{0b}=w_b>u_{ab}-\gk_{ab}$ for every $a>0$.

The communication device in the PCE uses a trade state and an absorbing rejection state. In the trade state, it recommends $b$ to the buyer and privately recommends each effort $a$ with probability $\zeta_a$ to the seller. The seller is prescribed effort $a$ after $b$ and shirking after every other trade. A good output leaves the state unchanged. After recommendation $a$ and a bad output, the device selects the rejection state with probability
\begin{align*}
\frac{(1-\gd)d_{ab}}{\gd(\overline U-\overline\gk_{\mathrm{PCE}})},
\end{align*}
and otherwise leaves the state unchanged. This probability lies in $[0,1]$ and is therefore well-defined for sufficiently high $\gd$. Any buyer deviation leads to rejection from the next period, with shirking in an unexpected trade. The rejection state prescribes rejection and shirking in every unexpected trade.

The seller's state payoffs are $\overline U-\overline\gk_{\mathrm{PCE}}$ and zero, since the trade-state payoff equation reduces to $(1-\gd)\sum_a\zeta_a(u_{ab}-\gk_{ab})
+\gd(\overline U-\overline\gk_{\mathrm{PCE}})
=\overline U-\overline\gk_{\mathrm{PCE}}$. In the trade state, the seller's good-minus-bad continuation-payoff difference after recommendation $a$ is $(1-\gd)d_{ab}/\gd$, so that effort is optimal by \cref{lem:optimala}. The buyer is indifferent between trade $b$ and rejection; every other trade induces shirking and gives a negative payoff. In the rejection state and after a buyer deviation, shirking is optimal because future play is permanent rejection. These comparisons hold after every mediator history, and the one-shot deviation principle rules out every seller deviation. Thus the construction is a PCE.

\paragraph{Step 3.} Fix $(U,V)\in\operatorname{int}\mathcal F$ with $0<U<\overline U-\overline\gk_{\mathrm{PCE}}$. Choose $\widetilde V>V$ with $(U,\widetilde V)\in\operatorname{int}\mathcal F$. As in Step 3 of the proof of \cref{prop:Flbenchmark}, I construct a PCE in which the seller's payoff is $U$ and the average buyer's payoff converges to $\widetilde V$ as $\delta \ra 1$. I then use public randomization between this PCE and the PCE attaining $(U,0)$ to achieve a PCE that attains $(U,V)$.

The PCE has a main phase and a residual phase, beginning with a main phase. By the definition of $\mathcal F$, choose probabilities $\mu_0$ for rejection and $\mu_{ab}$ for trade $b>0$ and action $a\in A$ in that trade, with $\mu_0+\sum_{b>0}\sum_{a\in A}\mu_{ab}=1$, such that
\begin{align*}
\sum_{b>0}\sum_{a\in A}\mu_{ab}(u_{ab},v_{ab})
&=(U,\widetilde V),\\
\sum_{a\in A}\mu_{ab}v_{ab}
&\ge0,\qquad b=1,\dots,\overline b.
\end{align*}
Let $h_t$ denote the mediator history before period $t$'s recommendations and $U'_t$ the mediator's promised utility, or simply promise, to the seller at that history, with $U'_0=U$.\footnote{Here I refer to $U'_t$ as promised utility rather than continuation payoff because the seller does not (directly) observe the mediator's history, unlike in the no-mediation benchmark.} In each main-phase period, the device draws $(b,a)$ with probability $\mu_{ab}$ or rejection with probability $\mu_0$. It tells the buyer only her recommended action and prescribes effort $a$ in trade $b$ and shirking in every other trade to the seller. After recommendation $(b,a)$, use the promise update in \eqref{eq:y}; after prescribed rejection, set $U'_{t+1}=U'_t/\gd$. A buyer deviation induces shirking if trade occurs and permanent rejection from the next period. The update gives a good-minus-bad continuation-payoff difference of $(1-\gd)d_{ab}/\gd$, so every recommended effort is optimal by \cref{lem:optimala}. The recommendation mixtures satisfy the buyer's obedience constraints \eqref{eq:recursiveIC2}, and the seller's promise-keeping condition holds: $U'_t = (1-\gd) \bE[u_t|h_t] + \gd \bE[U'_{t+1} | h_t]$.

The main phase ends when the promise leaves a fixed interval around $U$. Choose $\eta>0$ with $2\eta<\min\{U,\overline U-\overline\gk_{\mathrm{PCE}}-U\}$, and continue while $U'_t\in[U-\eta,U+\eta]$. Let $R:=\max_{b>0}w_b+\max_{b>0}d_{\overline a b}$. As in the proof of \cref{prop:Flbenchmark}, every promise update satisfies $|U'_{t+1}-U'_t|\le(1-\gd)R/\gd<\eta$ for sufficiently high $\gd$. Every next-period promise therefore lies in $[U-2\eta,U+2\eta]\subset(0,\overline U-\overline\gk_{\mathrm{PCE}})$. Once the promise exits the interval, the PCE begins the residual phase with a continuation PCE attaining $(U'_{t+1},0)$ from Step 2.


The residual phase has vanishing expected discounted weight. Let $\tau$ be its first period. Before exit, because $U'_t=(1-\gd)\bE[u_t\mid h_t]+\gd\bE[U'_{t+1}\mid h_t]$, \eqref{eq:identity}--\eqref{eq:maximal} apply with mediator histories in place of public histories and give $\bE[\gd^\tau]\le R \sqrt{1-\gd}/\eta \ra 0$. Each main-phase period gives the buyer expected payoff $\widetilde V$ conditional on $h_t$, and the residual phase gives zero. The initial payoff vector is therefore $(U,V_\gd)$, where $V_\gd:=\widetilde V(1-\bE[\gd^\tau])\ra\widetilde V$. For sufficiently high $\gd$, $V_\gd>V$. Public randomization between this PCE and the PCE attaining $(U,0)$, with probability $V/V_\gd$ on the former, attains $(U,V)$ exactly. Finally, Hausdorff convergence follows from the same compactness argument as in the proof of \cref{prop:Flbenchmark}.

\end{appendices}

\addtocontents{toc}{\protect\setcounter{tocdepth}{1}}
\renewcommand\bibname{References}

\pagebreak

{
\bibliographystyle{chicago}
\bibliography{references}
}

\pagebreak
\clearpage

\setcounter{section}{0}
\renewcommand{\thesection}{\Roman{section}}
\renewcommand{\theHsection}{supplementary.\Roman{section}}


\end{document}

%% file: dstyle2.tex
\usepackage[english]{babel}

\usepackage[nottoc,notlot,notlof]{tocbibind}
\usepackage{amssymb,mathrsfs,amsfonts}
\usepackage{amsmath, amsthm, xpatch, dsfont,ushort,sgamevar, float, framed, multirow, array, makecell, comment, url, graphicx, subcaption, mathtools, fnpct, setspace, pifont, bm, bbm,accents}
\usepackage{nicefrac, xfrac}
\usepackage{epstopdf}
\usepackage[normalem]{ulem}

\usepackage{lmodern}

\usepackage[final, babel, activate={true,nocompatibility}]{microtype}
\DeclarePairedDelimiterX{\infdivx}[2]{(}{)}{%
#1\;\delimsize\|\;#2%
}

\usepackage{lipsum}
\usepackage[utf8]{inputenc} 
\usepackage[T1]{fontenc} 
\usepackage{titlesec}
\usepackage{thmtools}

\usepackage[round,comma,authoryear,longnamesfirst,sort&compress]{natbib}
\usepackage[bottom, multiple]{footmisc}
\usepackage[top=1.5in, bottom=1.5in, left=1.25in, right=1.25in]{geometry}
\usepackage{xcolor}
\definecolor{navyblue}{rgb}{0.0, 0.13, 0.5}
\definecolor{amaranth}{rgb}{0.9, 0.17, 0.31}
\definecolor{brightpink}{rgb}{1.0, 0.0, 0.5}
\definecolor{cgreen}{rgb}{0.0, 0.42, 0.24}
\definecolor{cobalt}{rgb}{0.0, 0.28, 0.67}
\definecolor{coquelicot}{rgb}{1.0, 0.22, 0.0}
\definecolor{deepcarrotorange}{rgb}{0.91, 0.41, 0.17}
\definecolor{portlandorange}{rgb}{1.0, 0.35, 0.21}
\definecolor{burntorange}{rgb}{0.8, 0.33, 0.0}
\definecolor{dukeblue}{rgb}{0.0, 0.0, 0.61}
\definecolor{brickred}{rgb}{0.8, 0.25, 0.33}
\definecolor{brown(web)}{rgb}{0.65, 0.16, 0.16}
\definecolor{ceruleanblue}{rgb}{0.16, 0.32, 0.75}
\definecolor{sgblue}{rgb}{0.0510, 0.2039, 0.4706}
\definecolor{sgbrown}{rgb}{0.5020, 0.2510, 0.0000}

\usepackage{hyperref}
\hypersetup{
colorlinks=true,
linkcolor=black,
filecolor=black,
citecolor=black,
urlcolor=blue,
}
\usepackage[capitalize,noabbrev,nameinlink]{cleveref}

\crefformat{onlineappendix}{#2Supplementary Appendix~#1#3}
\Crefformat{onlineappendix}{#2Supplementary Appendix~#1#3}

\DeclareMathAlphabet{\mathsf}{OT1}{\sfdefault}{m}{n}
\SetMathAlphabet{\mathsf}{bold}{OT1}{\sfdefault}{b}{n}

\usepackage{enumitem}

\renewcommand\thmcontinues[1]{Continued}

\newtheorem{claim*}{Claim}[]

\usepackage{tocloft}

\newtheorem{prop*}{Proposition}[]
\newtheorem{bprop*}{Proposition}
\newtheorem{bthm*}{Theorem}
\newtheorem{thm*}{Theorem}[]

\newtheorem{ax*}{Axiom}[]
\newtheorem{lem*}{Lemma}[]
\newtheorem{rem*}{Remark}[]
\newtheorem{asp*}{Assumption}[]
\newtheorem{cor*}{Corollary}[]
\newtheorem{pty*}{Property}[]
\newtheorem{def*}{Definition}[]
\newtheorem{cond*}{Condition}[]

\newtheorem*{propnonumber*}{Proposition}

\usepackage{caption}
\usepackage[labelfont=bf]{caption}

\usepackage{tikz}
\usepackage{pgfplots}
\usetikzlibrary{automata,positioning}
\usetikzlibrary{arrows,intersections}

\newtheorem{examplex}{Example}
\newenvironment{eg*}
{\pushQED{\qed}\examplex}
{\popQED\endexamplex}

\newcommand{\ra}{\rightarrow}

\newcommand{\ve}{\varepsilon}

\newcommand{\bE}{\mathbf{E}}
\newcommand{\bR}{\mathbf{R}}

\newcommand{\gl}{\lambda}
\newcommand{\ga}{\alpha}

\newcommand{\gd}{\delta}
\newcommand{\gs}{\sigma}

\newcommand{\gk}{\kappa}
\newcommand{\egg}{e.g.,~}

\usepackage[title,titletoc,toc]{appendix}

\renewcommand{\bibname}{References}

\renewcommand{\texttt}[1]{%
\begingroup
\ttfamily
\begingroup\lccode`~=`/\lowercase{\endgroup\def~}{/\discretionary{}{}{}}%
\begingroup\lccode`~=`[\lowercase{\endgroup\def~}{[\discretionary{}{}{}}%
\begingroup\lccode`~=`.\lowercase{\endgroup\def~}{.\discretionary{}{}{}}%
\catcode`/=\active\catcode`[=\active\catcode`.=\active
\scantokens{#1\noexpand}%
\endgroup
}

\makeatletter
\renewenvironment{proof}[1]
[\proofname]{\par\pushQED{\qed}\normalfont%
\topsep6\p@\@plus6\p@\relax
\trivlist\item[\hskip\labelsep\bfseries#1\@addpunct{.}]%
\ignorespaces}{%
\popQED\endtrivlist\@endpefalse
}
\makeatother

\pgfplotsset{compat=1.14}